\documentclass[iop]{emulateapj}
\usepackage{amssymb}
\usepackage{natbib}
\usepackage{amsmath}
\usepackage{subfigure}
\usepackage{footnote}
\usepackage{color}
\usepackage{mdwlist}
\usepackage{enumerate}
\usepackage{graphicx}
\usepackage[bottom]{footmisc}

\newcommand{\fexii}{{\ion{Fe}{12}}}

\newcommand{\fexxiv}{{\ion{Fe}{24}}}
\newcommand{\fexxiii}{{\ion{Fe}{23}}}

\newcommand{\kps}{{km\,s$^{-1}$}}
\def\arcsec{$^{\prime\prime}$}
          
\begin{document}
\title{Broad Non-Gaussian \fexxiv~Line Profiles in the Impulsive Phase \\ of the 2017 September 10 X8.3--class Flare Observed by \emph{Hinode}/EIS}
\author{Vanessa Polito}
\affil{Harvard-Smithsonian Center for Astrophysics, 60 Garden Street, Cambridge MA 01238, USA}
\author{Jaroslav Dud{\'{\i}}k} 
\author{Jana Ka{\v s}parov{\'a}}
\author{Elena Dzif{\v c}{\'a}kov{\'a}}
\affil{Astronomical Institute of the Czech Academy of Sciences, Fri\v{c}ova 298, 251 65 Ondrejov, Czech Republic}
\author{Katharine K. Reeves}
\author{Paola Testa}
\affil{Harvard-Smithsonian Center for Astrophysics, 60 Garden Street, Cambridge MA 01238, USA}
\author{Bin Chen}
\affil{Center for Solar-Terrestrial Research, New Jersey Institute of Technology, Newark, NJ, United States}
\begin{abstract}
We analyze the spectra of high temperature \fexxiv~lines observed by \emph{Hinode}/EIS during the impulsive phase of the X8.3--class flare on September 10, 2017. The line profiles are broad, show pronounced wings, and clearly depart from a single Gaussian shape. The lines can be well fitted with $\kappa$ distributions, with values of $\kappa$ varying between~$\approx$1.7 to 3. The regions where we observe the non-Gaussian profiles coincide with the location of high-energy ($\approx$100--300 keV) HXR sources observed by \emph{RHESSI}, suggesting the presence of particle acceleration or turbulence, also confirmed by the observations of a non-thermal microwave sources with the \emph{Expanded Owens Valley Solar Array} (EOVSA) at and above the HXR looptop source. We also investigate the effect of taking into account $\kappa$ distributions in the temperature diagnostics based on the ratio of the \fexxiii~263.76~\AA~and \fexxiv~255.1~\AA~EIS lines. We found that these lines can be formed at much higher temperatures than expected (up to log($T$\,[K])\,$\approx$\,7.8), if departures from Maxwellian distributions are taken into account. Although larger line widths are expected because of these higher formation temperatures, the observed line widths still imply non-thermal broadening in excess of ~200\,\kps. The non-thermal broadening related to HXR emission is better interpreted by turbulence rather than chromospheric evaporation.

\end{abstract}

\section{Introduction}
\label{Sect:intro}

Solar flares \citep[e.g.,][]{Fletcher11,Schmieder15} are bright manifestations of the release of magnetic energy via the process of magnetic reconnection, leading to plethora of observed dynamics. From the viewpoint of spectroscopy, one of the long-standing problems is the presence of broad profiles of flare lines, usually interpreted in terms of non-thermal (turbulent) velocities \citep[e.g.,][]{Doschek79,Doschek80X,Feldman80M,Culhane81,Antonucci82,Antonucci86,Antonucci89,Tanaka82,Landi03,DelZanna06,Milligan11,Brosius13,Young13,Tian14,Polito15,Polito16,Bamba17,Lee17,Woods17}. Typically, the largest non-thermal velocities, of the order of 100--200\,\kps, are observed in the hottest flare lines available, such as \ion{Fe}{21}--\ion{Fe}{26}. Such high non-thermal broadenings occur exclusively during the start or the impulsive phases of a flare, followed by a decrease to about 60\,\kps~during the gradual phase. Reports of much smaller velocities of $\approx$40\,\kps~in the impulsive phase have been also made \citep[][Table 2 therein]{Young15} using the \ion{Fe}{21} line observed by the \emph{IRIS} satellite \citep{DePontieu14}. Progressive broadenings, from about 40 to 90\,\kps~have been also reported during the rise phase of a long-duration X1-class flare \citep[see, e.g., Sect. 3.2 and Fig. 10 of][]{Dudik16}. If the broad flare line profiles are interpreted in terms of the equivalent (Doppler) ion temperatures, values above 100\,MK can be obtained \citep[e.g.,][]{Antonucci86,Antonucci89}, which are almost an order of magnitude higher than the corresponding electron temperatures of several 10$^7$\,K at which the hot flare lines are formed.

For the sake of completeness, we note that non-thermal broadening of EUV lines of the order of several tens of \kps~are found also outside flaring regions, such as in the active region transition region and corona \citep[e.g.,][]{DePontieu15,Brooks16,Testa16}. Further, large non-thermal velocities are not found exclusively in flares, but can also be found in microflares, where they can also reach 145\,\kps~\citep[][Table 2 therein]{Brosius13}.


Recently, \citet{Jeffrey16,Jeffrey17} showed that the shape of the broad line profiles during flares can deviate from a single Gaussian. In particular, non-Gaussian wings were observed in the \ion{Fe}{16} and \ion{Fe}{23} lines with the Extreme-Ultraviolet Imaging Spectrometer \citep[EIS,][]{Culhane07} onboard the \emph{Hinode} spacecraft \citep{Kosugi07}. These line profiles were fitted by the non-Maxwellian $\kappa$-distributions \citep[][]{Olbert68,Vasyliunas68a,Vasyliunas68b,Livadiotis09,Livadiotis17,Dudik17b}, yielding low values of $\kappa$ ($\approx$~2--3), which indicate strong departures from the Gaussian shape. We note that similar findings were obtained for active region brightenings in transition-region lines \citep{Dudik17a}, as well as coronal holes \citep{Jeffrey18}. \citet{Jeffrey17} found that the flare line profiles described by a $\kappa$-distribution could occur either due to ion acceleration or presence of turbulence. This is not surprising, since turbulence leads naturally to enhanced high-energy tails of the particle distributions, if the turbulent diffusion coefficient is inversely proportional to velocity \citep{Hasegawa85,Laming07,Che14}. In flares, such situation can lead to a $\kappa$-distribution \citep{Bian14}. The high-energy tails can, however, be also produced by particle acceleration \citep[e.g.,][]{Gordovskyy13,Gordovskyy14,Gontikakis13} or wave-particle interaction involving whistler waves, as shown by \citet{Vocks08,Vocks16}.

The $\kappa$-distributions of electrons have been suggested in flares using the \emph{RHESSI} \citep{Lin02} spacecraft. The method consisted of fitting the X-ray bremsstrahlung spectra, which allows to determine the $\kappa$ index for the high-energy tail \citep{Oka13,Oka15} or the entire spectrum \citep{kasparova2009}. Alternatively, the $\kappa$ parameter can be determined by combining observations from the \emph{RHESSI} spacecraft and the Atmospheric Imaging Assembly \citep[AIA,][]{Lemen12,Boerner12} onboard the \emph{Solar Dynamics Observatory} \citep[\emph{SDO}, ][]{Pesnell12} using the mean electron flux spectrum \citep{Battaglia13,Battaglia15}. Yet another method involves the ratios of emission line intensities which are formed over a wide range of energies of the impacting electrons that produce ionization and excitation. Such method have been used to detect strongly non-Maxwellian $\kappa$-distributions in the rise and impulsive phase of the X5.4--class solar flare of 2012 March 07 \citep{Dzifcakova18}.


In this work, we report on detection of strongly non-Gaussian and broad line profiles of \fexxiv~at the top of the flare loops during the strongest solar flare of the Solar cycle 24, i.e., the X8.3--class Flare of 2017 September 10. This is a limb flare, allowing for localizing the non-Gaussian profiles from a different viewpoint than the on-disk flares of \citet{Jeffrey16,Jeffrey17}. The paper is organized as follows. Section \ref{Sect:obs} describes the spectroscopic observations used and their reduction. Fitting of the line profiles is detailed in Sect. \ref{Sect:fit} and the results on temperature diagnostics based on the \fexxiv~255.10/\fexxiii~263.76~\AA~ratio is discussed in Sect.~\ref{Sect:temp_diagnostics}.
Section \ref{Sect:rhessi} describes the analysis of the \emph{RHESSI} spectra and fitting of the various spectral components. The results obtained are discussed in Sect. \ref{Sect:discussion} and a summary is given in Sect. \ref{Sect:Summary}.

\begin{figure}
\centering
\includegraphics[width=0.45\textwidth]{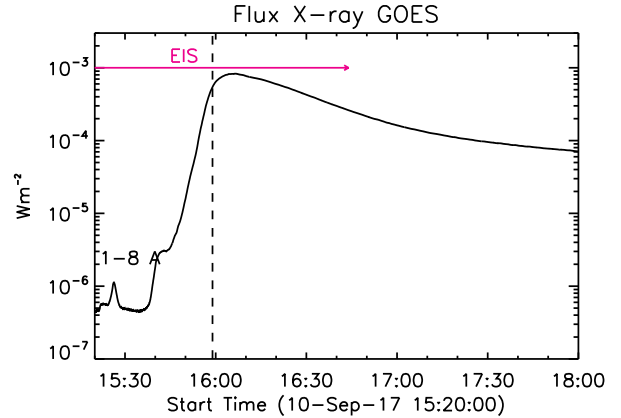}
\caption{\emph{GOES} soft X-ray light curve in the 1--8~\AA~channel for the X-class flare on September 10, 2017. The pink arrow indicates the duration of the EIS observation under study (which starts at 05:44~UT). The horizontal dotted line shows the time where we observe the non-Gaussian \fexxiv~line profiles. }
\label{Fig:goes}
\end{figure}

\begin{figure*}
\centering
\includegraphics[width=\textwidth]{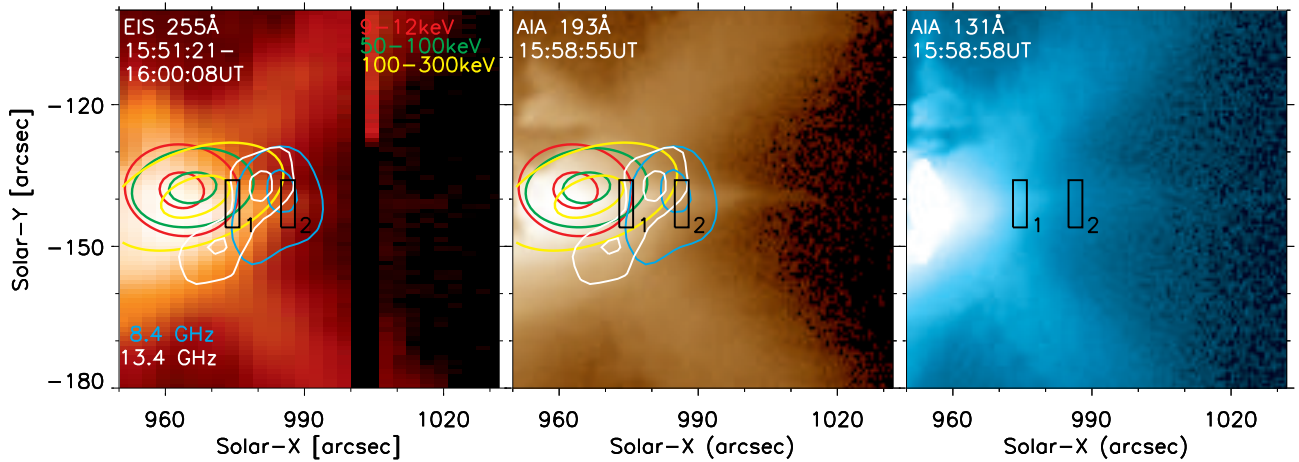}
\caption{Overview of the 10th September 2017 flare observation with \emph{Hinode}/EIS and \emph{SDO}/AIA. \emph{Left panel:} Intensity image formed in the \fexxiv~255.10~\AA~line during one of the EIS rasters (from 15:51:21 to 16:00:08~UT), obtained by performing a single Gaussian fit at each raster pixel. \emph{Middle and right panels:} AIA 193~\AA~and 131~\AA~images at the closest time to the EIS raster exposures number 8--9. The colored contours in the left and middle panels show the 50 and 90\%~of the maximum intensity of the HXR sources observed by \emph{RHESSI} during the time interval 15:58:40--15:59:40~UT, and the  microwave sources observed by \emph{EOVSA} between 15:59:08~UT and 15:59:12~UT, with different colors indicating different energy intervals as shown in the legend.
Finally, the black boxes 1 and 2 indicate the location where we perform the $\kappa$ fitting of the \fexxiv~255.10~\AA~(raster exposures 8--9) and 192.03~\AA~(raster exposures 12--13) lines respectively. }
\label{Fig:overview}
\end{figure*}

\section{Observations}
\label{Sect:obs}
The X8.3 flare under study occurred on September 10, 2017 in the AR~$\beta\gamma$~NOAA~12673 on the west limb of the Sun. The flare started at $\approx$15:40~UT and peaked at about 16:06:28~UT, as observed by the \emph{GOES} satellite in the 1--8~\AA~filter (see Fig.~\ref{Fig:goes}). Several authors have focused on the analysis of this event, which was observed simultaneously by different satellites, including \emph{SDO}, \emph{Hinode}, \emph{IRIS}, \emph{RHESSI}, and \emph{Fermi} \citep[e.g.,][]{Warren18,Long18,Seaton18,Yan18,Li18}. The flare was initiated by a fast eruption of a flux rope cavity starting at about~15:54 UT. It was also associated with increase in non-thermal electron energy flux as measured by \emph{Fermi} (see Fig. 1 in Long et al. 2018). Spectroscopic observations performed by \emph{Hinode}/EIS were studied in \citet{Warren18} and \citet{Li18}, who focused on studying the evolution of the hot plasma in a plasma sheet formed after the fast eruption and above the flare loops, from Solar-X coordinates~$\approx$960\arcsec~towards larger solar radii (see Fig.~\ref{Fig:overview}). 

In this work, we focus on analyzing the profiles of the high temperature lines observed by \emph{Hinode}/EIS (\fexxiv~and \fexxiii) in the plasma sheet at around 15:59~UT, as indicated by the horizontal line in Fig.\ref{Fig:goes}. We note that \citet{Warren18} and \citet{Li18} focused on the EIS spectra observed after 16:09\,UT, i.e., towards the peak and gradual phases of the flare. At these times, the \fexxiv~and \fexxiii~lines could be well-fitted by Gaussians (see, e.g., Figure 2 in \citeauthor{Li18} \citeyear{Li18} and Figure 10 in \citeauthor{Warren18} \citeyear{Warren18}). \citet{Long18} studied the EIS spectra associated with the fast eruption at about 15:42 and 15:54\,UT (see their Figures 3 and 4). Our analysis is complementary to these ones both in time and as well as because of its different focus on the non-Gaussian shape of line profiles.

The analysis of the EIS spectra is complicated by some well-known technical issues (as summarized in the EIS wiki page\footnote{http://solarb.mssl.ucl.ac.uk/eiswiki/Wiki.jsp?page=EISAnalysisGuide}), some of which are briefly described in Sect.~\ref{Sect:eis_data}. 
We use high-resolution \emph{SDO}/AIA images (0.6\arcsec~per pixel) in the 193~\AA~filters, which provide an essential context for the spectroscopic observations and useful information on the plasma emission measure (see Sect.~\ref{Sect:temp_diagnostics}). The level 1.5 AIA data were processed using the \emph{SolarSoft} \citep[SSW; ][]{Freeland98} routine \texttt{aia\_prep.pro}, which corrects for instrumental pointing errors and co-aligns images from different filters on a common platescale. We also analyzed hard X-ray (hereafter, HXR) images from the \emph{RHESSI} satellite, whose data reduction is discussed in Sect.~\ref{Sect:rhessi}. The flare was also observed in microwave wavelengths in 2.5--18 GHz by the Expanded Owens Valley Solar Array (\emph{EOVSA}). An overview of the \emph{EOVSA} observations is provided by \cite{Gary18}.

%
\subsection{EIS data reduction and instrumental issues}
\label{Sect:eis_data}
On September 10, 2017 EIS observed the AR 12673 while running a large 80-step raster from 05:44~UT to 16:44~UT, covering all the impulsive and part of the gradual phase of the X8.3--class flare (see Fig.~\ref{Fig:goes}). For each raster, the 80 2\arcsec~slit positions were separated by a 1\arcsec~jump, resulting in a 3\arcsec~raster step and field-of-view for the spectrometer of 240\arcsec~x~304\arcsec. The exposure time was 5 s, and the total duration of each raster was $\approx$535~s. The EIS study consisted of 15 spectral windows, which included 3 high temperature lines that are only observed during flares: \fexxiii~263.76~\AA~(log($T$\,[K])\,$\approx$\,7.15), \fexxiv~192.03 and 255.10~\AA~(log($T$\,[K])\,$\approx$\,7.2). The \fexxiii~263.76~\AA~line is believed to be largely free of blends, whereas the \fexxiv~192.03~\AA~and 255.10~\AA~lines are contaminated with unidentified emission at 1~MK, as well as other lines including \ion{Fe}{11}, \ion{Fe}{12}, \ion{Fe}{17} \citep{DelZanna08,DelZanna12}, especially if the width of the \fexxiv~lines is large. Nevertheless, during flares, \fexxiv~appears to be the dominant contribution \citep{DelZanna08}. 

The level 0 EIS data were converted to level 1 datacubes ($\lambda$, X-pixels, Y-pixels) by using the \texttt{eis\_prep.pro} routine with some of the standard options \footnote{ftp://sohoftp.nascom.nasa.gov/solarsoft/hinode/eis/doc/eis\_notes/01\_EIS\_PREP/eis\_swnote\_01.pdf} , including the keyword \texttt{refill} to interpolate the missing pixels. We note that the interpolation is a necessary step because of the large number of missing pixels in this observation. According to the EIS Software Note 13, the interpolation works well since the EIS spatial resolution is 3--4 pixels in the Solar Y direction. This means that a signal within a given pixel contains a significant component from the neighboring spatial element. We also note that the interpolation of EIS spectra is a standard procedure before non-Gaussian fitting (see Section 4.7 of \citeauthor{Jeffrey16} \citeyear{Jeffrey16}, also \citet{Jeffrey17,Jeffrey18}).
However, in some cases our profiles still show 1 or 2 missing pixels which could not be interpolated by \texttt{eis\_prep.pro}. Finally, because of the uncertainty in the EIS radiometric calibration and its evolution \citep[see, e.g.][]{DelZanna13,Warren14}, we use the \texttt{/noabs} keyword in \texttt{eis\_prep.pro} and obtain the spectra in data number (DN).

We estimated an offset of about 16.5 pixels between the short-wavelength (SW, including the \fexxiv~192.03 \AA~line) and long-wavelength (LW, including the \fexxiv 255.10~\AA~line) CCD channels by using the routine \texttt{eis\_ccd\_offset.pro}. Although we took into account this offset when co-aligning spectra from the two CCDs, for simplicity the spectra are labelled with their original "uncorrected" pixel position. We also note that because of the offset between the two CCDs and the fact that the instrumental width varies with the CCD Y-pixel position, this latter will be different for lines which are included in different CCD channels, such as the \fexxiv~192.03~\AA~ and the 255.10~\AA~(or \fexxiii~263.76~\AA) line. This difference needs to be taken into account when fitting the line profiles, as discussed in Sect.~\ref{Sect:fit}. Further, we correct for the spectral tilt by using the SSW routine \texttt{eis\_slit\_tilt\_array.pro}. 

Finally, \cite{Warren18} pointed out the effects of the EIS assymmetric PSF, which causes apparent red and blueshifts in the centroid of the \fexxiv~line (and other bright lines) on either sides of the plasma sheet for the flare under study \citep[see Fig. 12 of ][]{Warren18}. We note that this instrumental effect does not influence the result of our analysis, as we observe \fexxiv~spectra with pronounced wings in several locations across the plasma sheet in the direction perpendicular to its length. In addition, if this effect was responsible for creation of the non-Gaussian wings, it should be seen in other conditions, for example in the same lines at different times. This is not the case, as these strong lines become Gaussian after 16:09\,UT \citep{Li18,Warren18}.

\begin{figure*}
\centering
\includegraphics[width=0.6\textwidth]{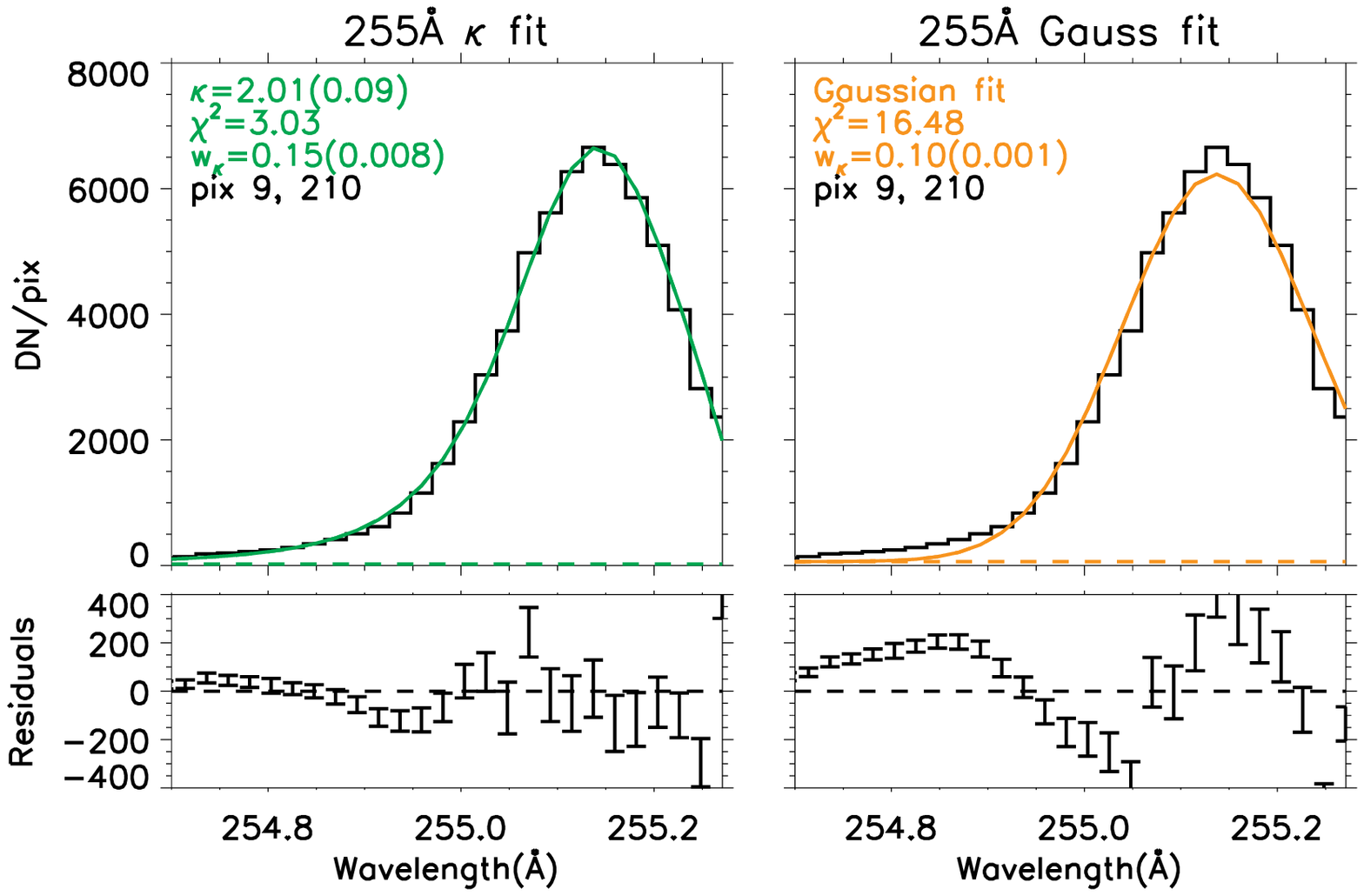}
\includegraphics[width=0.6\textwidth]{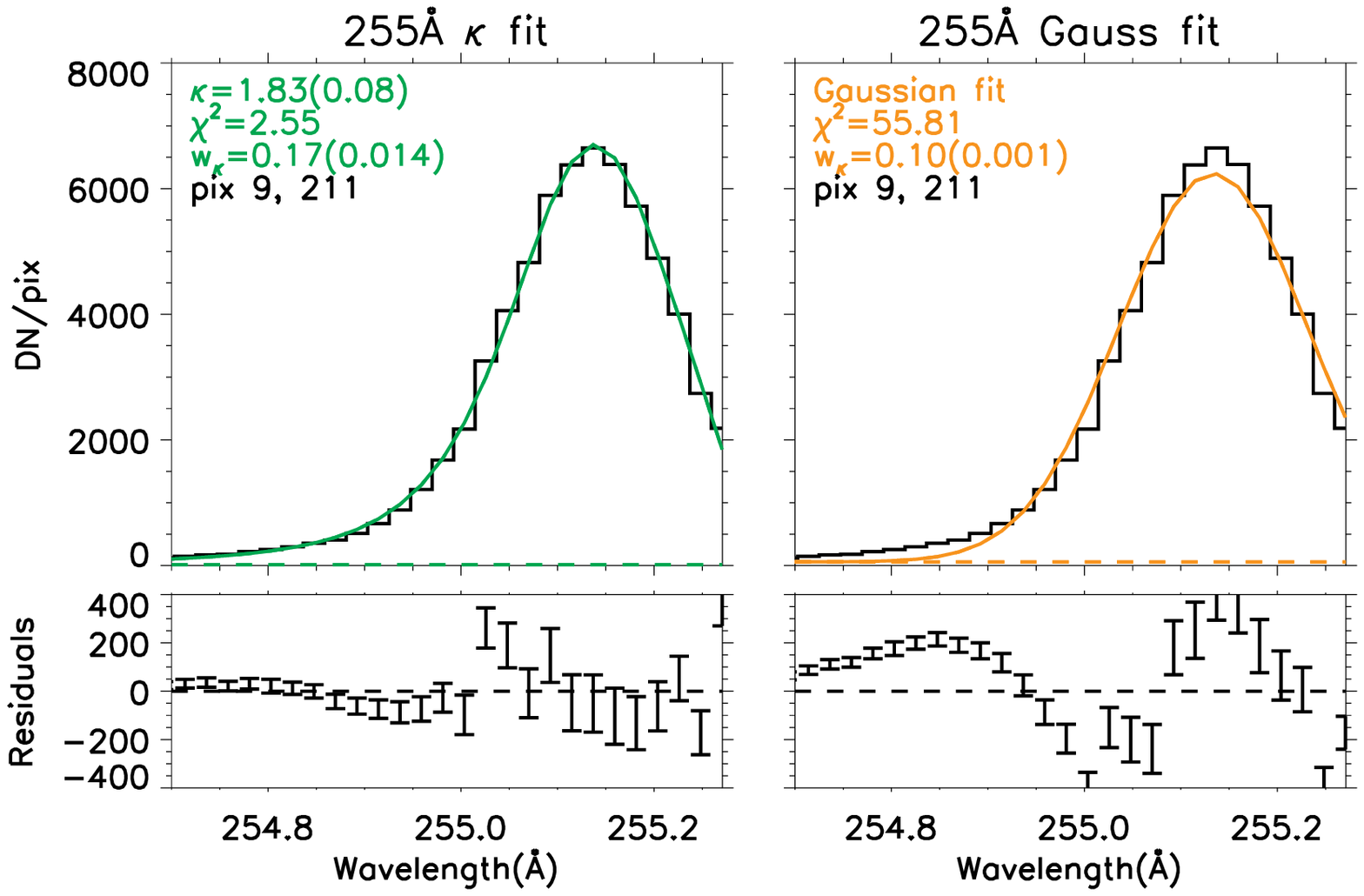}
\caption{Example of $\kappa$ (left panels) and single Gaussian (right panels) fits of the EIS \fexxiv~255.10~\AA~line during the raster exposure 9 and the slit pixels 210 and 211 (top and bottom panels respectively) inside box 1 of Fig.~\ref{Fig:overview}. For each fit, the value of $\chi^{2}$ is indicated in the left top part of the panel, as well as the width of the line obtained by the fit and its uncertainty. For the $\kappa$ fit, also the value of $\kappa$ and associated error are shown. Note that the line width has a different interpretation for a $\kappa$ and Gaussian fit, see text for more details. }
\label{Fig:fit_255_exp9}
\end{figure*}
\begin{figure*}
\centering
\includegraphics[width=0.8\textwidth]{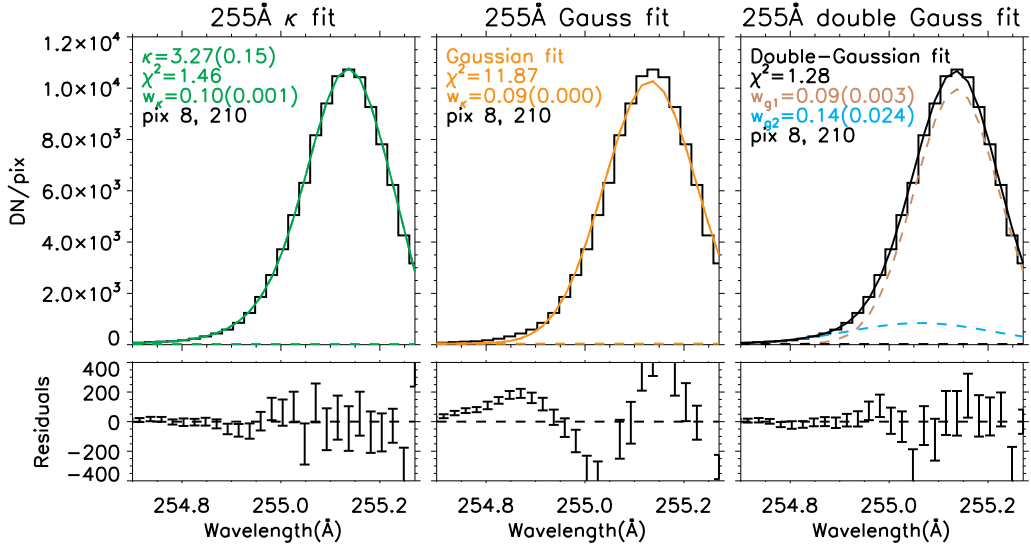}
\caption{Example of $\kappa$ (left panel) , single Gaussian (middle panel) and double-Gaussian (right panel) fits for the EIS \fexxiv~255.10~\AA~line during the raster exposure (X-pixel) 8 and the slit Y-pixel 210 (box\,1 of Fig.~\ref{Fig:overview}). See caption of Fig.~\ref{Fig:fit_255_exp9} and text for more details. }
\label{Fig:fit_255_exp8}
\end{figure*}

\begin{figure*}
\centering
\includegraphics[width=\textwidth]{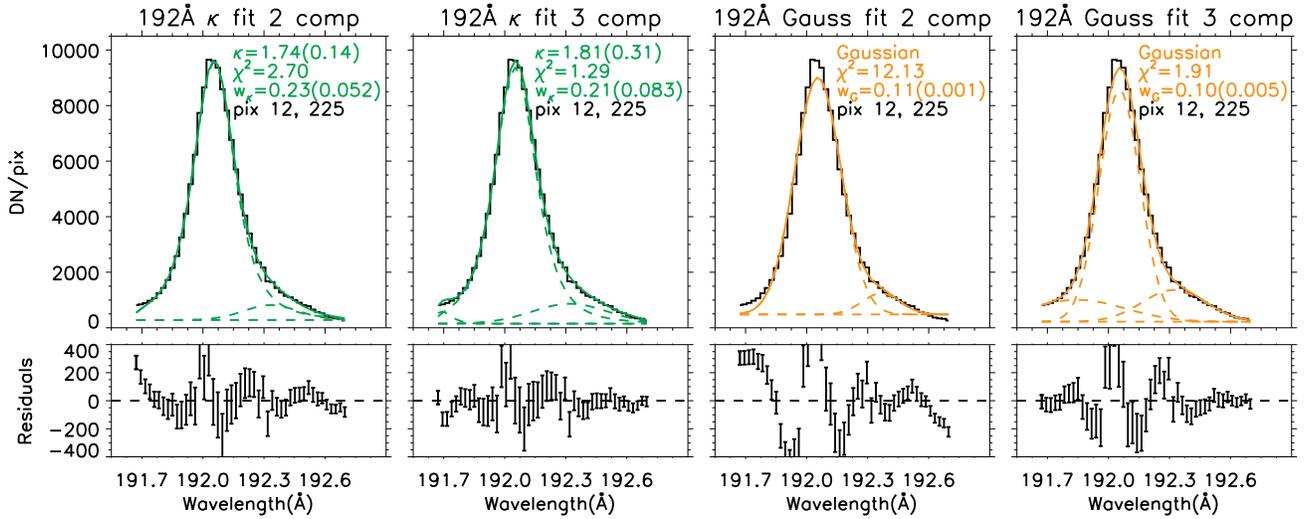}
\caption{\emph{From left:} Example of $\kappa$ fit and Gaussian fit with 1 or two blends for the EIS \fexxiv~192.03~\AA~line during the raster exposures 12 and the slit pixels 225 (box\,2 of Fig.~\ref{Fig:overview}). See caption of Fig.~\ref{Fig:fit_255_exp9} for more details on the figure.}
\label{Fig:fit_192}
\end{figure*}

\begin{figure*}
\centering
\includegraphics[width=0.6\textwidth]{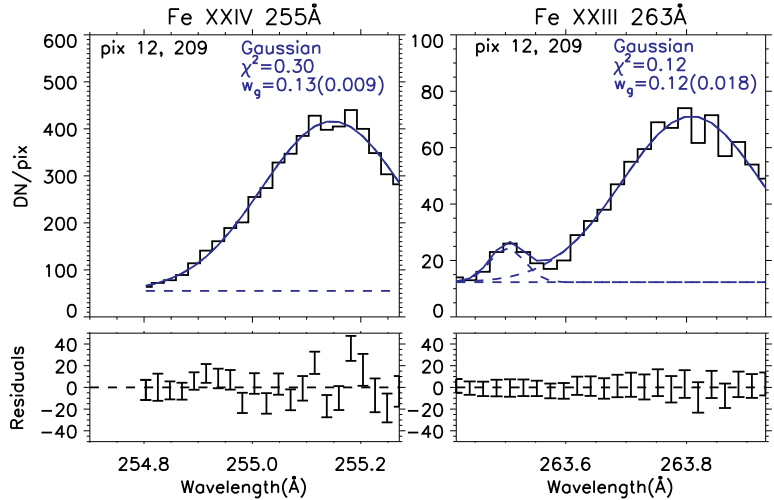}
\caption{Sample spectra of the  \fexxiv~255.10~\AA~(left panel) and \fexxiii~263.76~\AA~(right  panel) lines for raster exposure 12, Y-pixel 209 for the LW detector (corresponding to Y-pixel $\approx$225 in the SW detector). The spectra show that the lines are quite broad at this location, with $w_\mathrm{G}$ equal to 0.13$\pm$0.009~\AA~(or $FWHM_\mathrm{G}$ = 0.030$\pm$0.02~\AA) for the \fexxiv~255.10~\AA~line.}
\label{Fig:fe_23_24}
\end{figure*}

\begin{figure}
\centering
\includegraphics[width=0.4\textwidth]{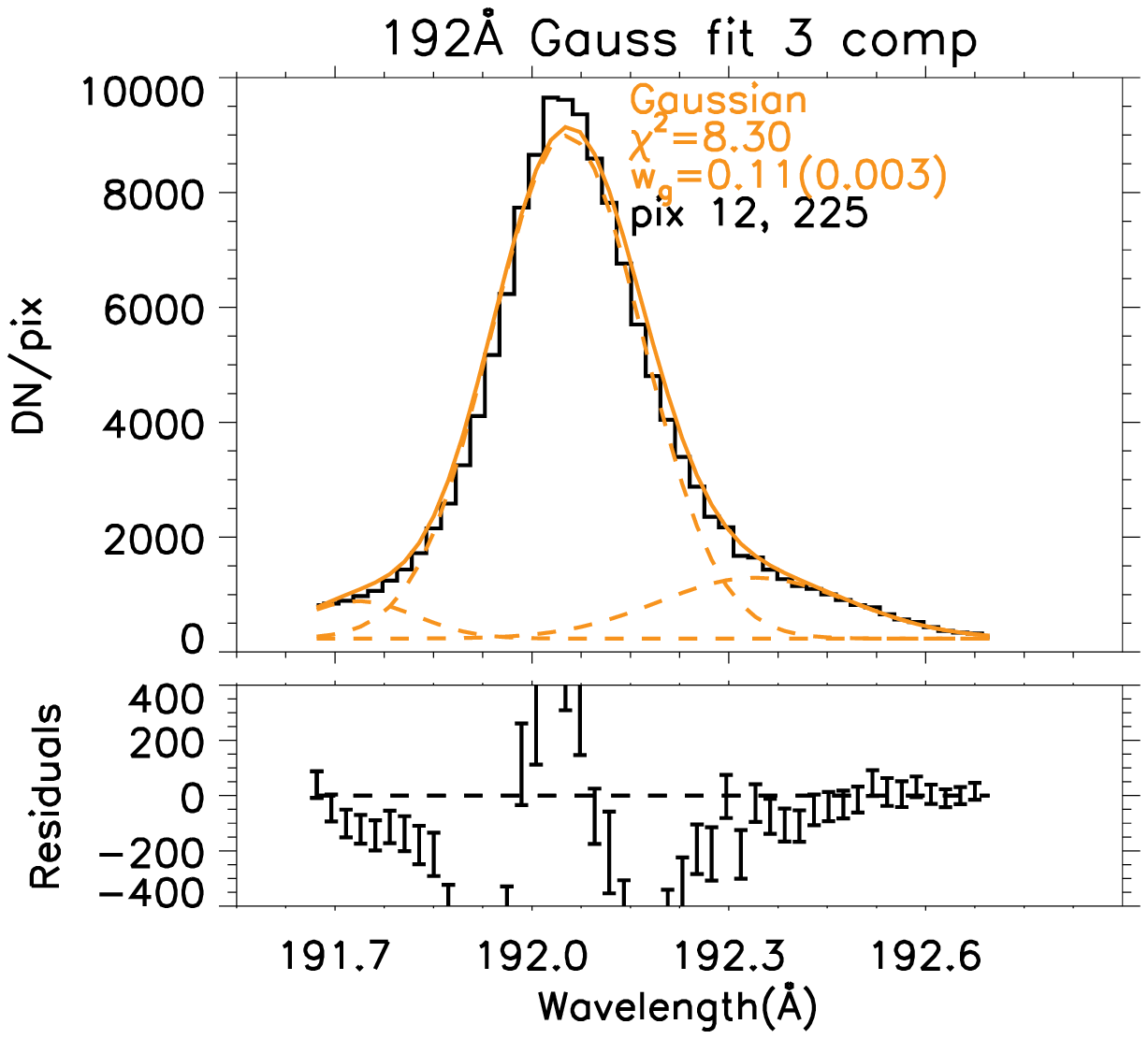}
\caption{Gaussian fit with two blends for the EIS \fexxiv~192.03~\AA~line during the raster exposure 12 and slit Y-pixels 225 (box\,2 of Fig.~\ref{Fig:overview}), obtained by fixing the minimum value of the line width to $w_\mathrm{G}$ of around 0.11~\AA~(or $FWHM_\mathrm{G}$ 0.026~\AA, see text for more details).}
\label{Fig:fit_192_width}
\end{figure}

\section{Fitting of the \fexxiv~and \fexxiii~lines}
\label{Sect:fit}
Mostly symmetric \fexxiv~line profiles with pronounced wings were observed at several locations along the plasma sheet feature during the EIS raster which was running between 15:51~UT and 16:00:08~UT. 
Figure~\ref{Fig:overview} (left panel) shows the intensity map of the EIS \fexxiv~255.10~\AA~line during this raster, which has been obtained by performing an approximate single-Gaussian fit at each pixel location. We note that the EIS slit rasters from right to left, and therefore the intensity map is actually a composite image obtained at different times. The profiles that we analyze in this work are observed in boxes 1 and 2 (shown in Figure~\ref{Fig:overview}), whose locations correspond respectively to the X-pixels 8--9 and 12--13 of the EIS datacube. Boxes 1 and 2 are located above the loop tops, at the bottom and along the plasma sheet structure respectively. The right panel shows the AIA~193~\AA~image at~$\approx$15:59~UT, which is the approximate time where the EIS slit was rastering the location indicated by box\,1. Overlaid on both images on left and right panels are the contours \emph{RHESSI} sources at different energy intervals, as indicated by the legend on the top right. The contour levels show the 50\% and 90\% of the maximum intensity of the \emph{RHESSI} images, which are integrated over the interval 15:58:40--15:59:40~UT by using the standard image reconstruction algorithms (see Sect.~\ref{Sect:rhessi}). The \emph{RHESSI} sources show the presence of very energetic electrons (above 100 keV) and coincide with the loop top of the flare, suggesting that we are observing a coronal HXR source. Such a bright coronal source cannot be interpreted in the standard
thin-target scenario, but it shows properties similar to some of the coronal
hard X-ray sources discussed in \citep[e.g.,][ and references therein]{Krucker08a}. Note that there is also a \emph{RHESSI} footpoint source outside the field-of-view of Figure~\ref{Fig:overview}, which is detectable above $\approx$~30 keV, see \cite{Gary18}. Figure~\ref{Fig:overview} also shows a microwave source observed by \emph{EOVSA} between 15:59:08~UT and 15:59:12~UT. This source is located at and above the HXR looptop source, with a characteristic non-thermal gyrosynchrotron spectral shape and a maximum brightness temperature of over $3\times 10^9$ K, suggesting the presence of highly energetic, mildly relativistic electrons in the loop top region.

The \fexxiv~255.10~\AA~line is intense enough in the location of box\,1, so that the fitting of the spectra (in particular the line wings) can be reliably performed. This location coincides well with the contours of the \emph{RHESSI} high-energy sources. The line is saturated on the left side of box\,1, and too faint on its right side. On the other hand, the \fexxiv~192.03~\AA~is saturated where the \emph{RHESSI} sources are most intense, and we can reliably fit the line only in the region indicated by box\,2. Although the saturation threshold of EIS is around 1.5$\times$10$^4$DN, we believe that non-linear effects in the instrumental gain might be important below this value, resulting in an underestimation of the peak for very intense lines. For this reason, we do not analyze line profiles with peak intensities above $\approx$10$^{4}$ DN. We also only select profiles where the interpolation algorithm has removed most of the missing pixels, with the exception of maximum 2 pixels. 

To fit the \fexxiv~255.10~\AA~and 192.03~\AA~lines, we use either a single (or multi-) Gaussian or a $\kappa$ fit, this latter performed by using the method described in Section 2.3 of \citet{Dudik17a}. In particular, we perform a convolution of the EIS instrumental Gaussian profile and a $\kappa$ profile by using a modified version of the SSW \texttt{comp\_gauss.pro} routine. This convolution has the form \citep[cf.,][]{Jeffrey16,Jeffrey17}
\begin{equation}
I(\lambda)	= I_0 \int\limits_{-\infty}^{\infty} \mathrm{e}^{-\frac{(\lambda - \lambda')^2}{2w_\mathrm{instr}^2}} \left(1 + \frac{(\lambda' - \lambda_0)^2}{2(\kappa-3/2)w_\kappa^2} \right)^{-\kappa} \mathrm{d}\lambda'\,,
\label{Eq:convolution}
\end{equation}
where $I_0$ is the peak intensity, $\lambda_0$ is the wavelength of the line center, $w_\kappa$ is characteristic width, and $\kappa$ is the non-Maxwellian parameter.

As mentioned in Sect.~\ref{Sect:eis_data}, the EIS instrumental width $w_\mathrm{instr}$ varies between the two \fexxiv~lines and it is estimated to be $\approx$0.0704~\AA~and 0.0698~\AA~ for the 192.03~\AA~and 255.10~\AA~lines respectively, as calculated using the SSW routine \texttt{eis\_slit\_width.pro}. 

For both the Gaussian and $\kappa$ fits, the weights \emph{W} are given by: 
\begin{equation}
W=\sqrt{\frac{1}{\sigma^2(I(\lambda_i))}}
\end{equation}
where $\sigma(I(\lambda_i))$ are the intensity errors obtained from \texttt{eis\_prep.pro}, which take into account the photon statistics noise, pedestal, and the error of the dark current. 

To evaluate the goodness-of-fit, we use the reduced $\chi^2_\mathrm{red}$ \citep[e.g., Eq. (18) of][]{Dudik17a}, as well as the residuals \emph{R} of the fits, obtained as: 
\begin{equation}
R = I_\textrm{obs} (\lambda_i)- I_\textrm{fit}(\lambda_i)
\label{eq:2}
\end{equation}
where $I_\textrm{obs}$ and $I_\textrm{fit}$ are the observed spectra and fit respectively for each wavelength $\lambda_i$. 
Note that in the following sections we will mostly express the width of the fitted lines in terms of characteristic widths $w_\textrm{G}$ or  $w_\textrm{$\kappa$}$  as obtained from either the Gaussian or $\kappa$ fit respectively. In order to convert from $w_\textrm{$\kappa$}$ to FWHM$_\textrm{$\kappa$}$, we use Eq.~14 of \cite{Dudik17a}, 
which we recall here for the reader's convenience: 
\begin{equation}
w_\textrm{$\kappa$} =\frac{1}{8}\frac{FWHM_\mathrm{\kappa}}{(\kappa-3/2)(2^{1/\kappa}-1)}
\label{eq:kappa_width}
 \end{equation}

For the Gaussian fit, $w_\textrm{G}$ is simply equal to:
\begin{equation}
 w_\mathrm{G} = \frac{FWHM_\mathrm{G}}{\sqrt{8\mathrm{ln}2}}
\label{eq:gauss_width}
 \end{equation}
 
The fitting procedure for the \fexxiv~255.10~\AA~and 192.03~\AA~lines are discussed separately in Sects.~\ref{Sect:fe24_255} and \ref{Sect:fe24_192}.
\begin{figure*}
\centering
\includegraphics[width=0.8\textwidth]{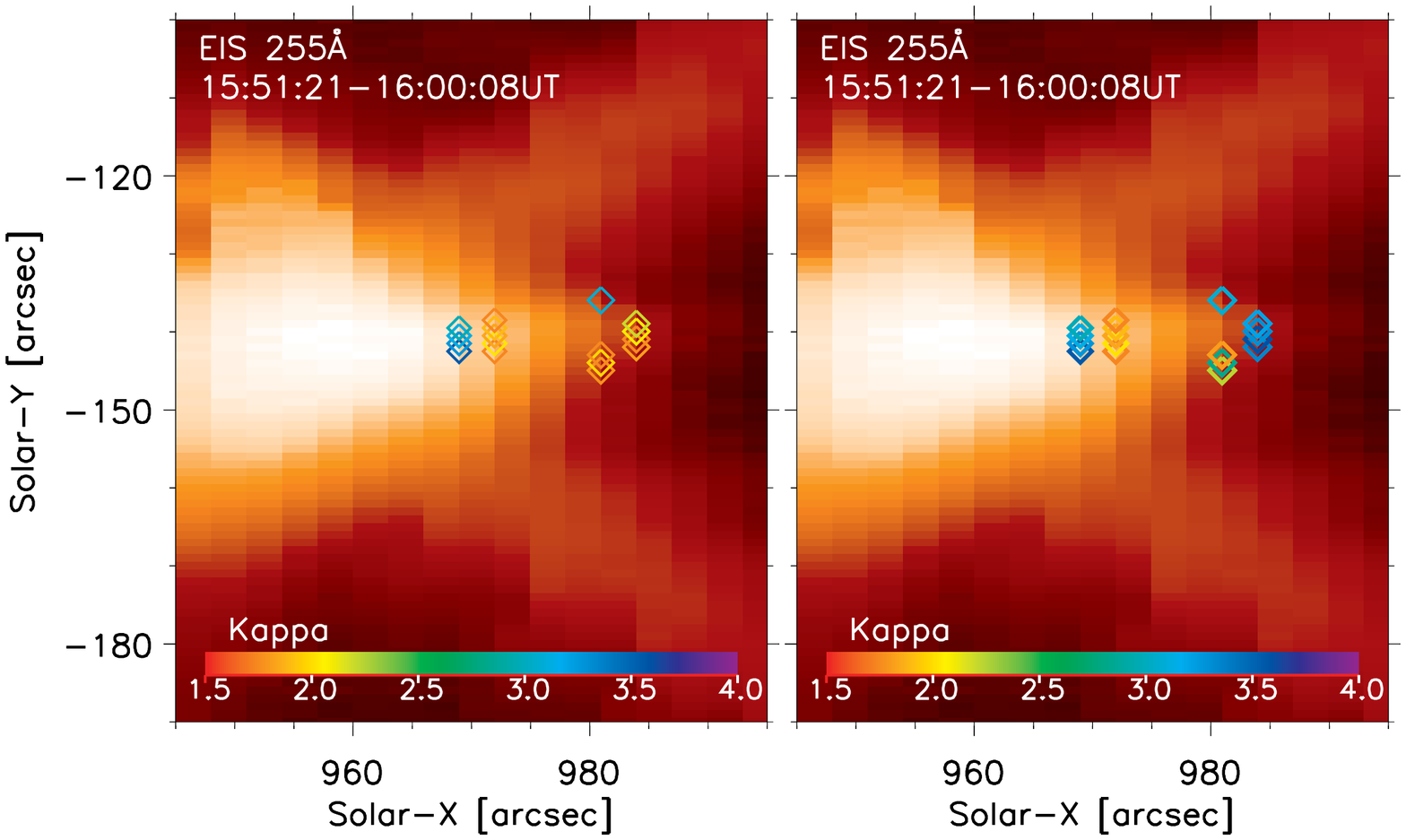}
\caption{Distribution of $\kappa$ values obtained by fitting the \fexxiv~255.10~\AA~and \fexxiv~192.03~\AA~lines overlaid on intensity images formed in the EIS \fexxiv~255~\AA~line}. The fit of the \fexxiv~192.03~\AA~line includes 1 or 2 blends (left and right panels respectively).
\label{Fig:kappa_map}
\end{figure*}

\subsection{\fexxiv~255~\AA~line fitting}
\label{Sect:fe24_255}
Figure~\ref{Fig:fit_255_exp9} shows two examples of \fexxiv~255.10~\AA~spectra at X-pixels = 9 and Y-pixels = 210 and 211 within box\,1, which have been fitted with a single $\kappa$ (left panel) and Gaussian (right panel) function. The results of the fit ($\chi^2_\mathrm{red}$ value, $\kappa$ value and line width expressed in \AA, with corresponding errors) are indicated in the same panels. At the bottom of both panels, we show the fitting residuals along the spectra (see Eq.~\ref{eq:2}).
We found that in both cases the single $\kappa$ fit performs much better than the Gaussian one, with considerably lower values of $\chi^2_\mathrm{red}$ and residuals. The same result is obtained for all the "good" pixels in box\,1 ($\approx$10) that we selected according to the criteria described above, as also discussed in Sect.~\ref{Sect:discussion}. 
We note that the red wing of the line is partially outside the spectral window, whereas the blue wing can be properly fitted. The fact that part of the spectrum is missing can potentially create some problems in the convolution part of the fitting algorithm. To rule this out, we compared the results of the convolution (Eq. \ref{Eq:convolution}) and fit with a purely $\kappa$-profile \citep[][]{Dudik17a},
\begin{equation}
I(\lambda)	= I_0 \left(1 + \frac{(\lambda - \lambda_0)^2}{2(\kappa-3/2)w_\kappa^2} \right)^{-\kappa} \mathrm{d}\lambda\,,
\label{Eq:kappa}
\end{equation}
and verified that there is no significant difference (within the uncertainties) between the $\kappa$ and line width values estimated using the convolution or purely $\kappa$ fit. This is not surprising, since the lines are much broader than the EIS instrumental width. 
As discussed in \cite{Dudik17a}, the pronounced wings can also be reproduced by a combination of two Gaussians profiles, which can be interpreted as the superposition of emission from plasma formed at two different temperatures. To investigate this possibility, we performed a double-Gaussian fit of the \fexxiv~255.10~\AA~in the same pixels of box\,1. Since the spectrum is missing its red wing, in most cases the uncertainties of the two-component fits were too high (with a $\chi^2_\mathrm{red}$ value much less than 1), except from one pixel (raster exposure 8, Y-pixel 210). Figure~\ref{Fig:fit_255_exp8} shows the comparison of the $\kappa$, Gaussian and double-Gaussian fit for this single case. 
The FWHM obtained from the double-Gaussian fit (using Eq.~\ref{eq:gauss_width}) are of the order of 0.20 and 0.34~\AA~for the two Gaussian components, which can be used to estimate the required non-thermal velocity as follows: 
\begin{equation}
v_\textrm{nth}= \frac{c}{\lambda_0}\frac{1}{2\sqrt{ln2}}\sqrt{\left(w^2-w_\textrm{instr}^2-w_\textrm{th}^2 \right)}\,.
\label{eq:nonth}
\end{equation}

Using the equation above, we can estimate that the line width of the broader Gaussian component corresponds to very large non-thermal velocities (of the order of $\approx$300 km\,s$^{-1}$). Although we cannot completely rule out the possibility of fitting the large wings with two Gaussian components, a single $\kappa$ fit appears as a possible simple explanation to account for the non-Gaussian shape of the line. Note that the line width obtained from the $\kappa$ fit ($\approx$~0.10~\AA, see Fig.~\ref{Fig:fit_255_exp8}) corresponds to a FWHM$_{\kappa}$ of 0.19~\AA~(using Eq.~\ref{eq:kappa_width}) and thus to a lower value of non-thermal velocity ($\approx$150~km~s$^{-1}$).

\begin{figure*}
\centering
\includegraphics[width=0.48\textwidth]{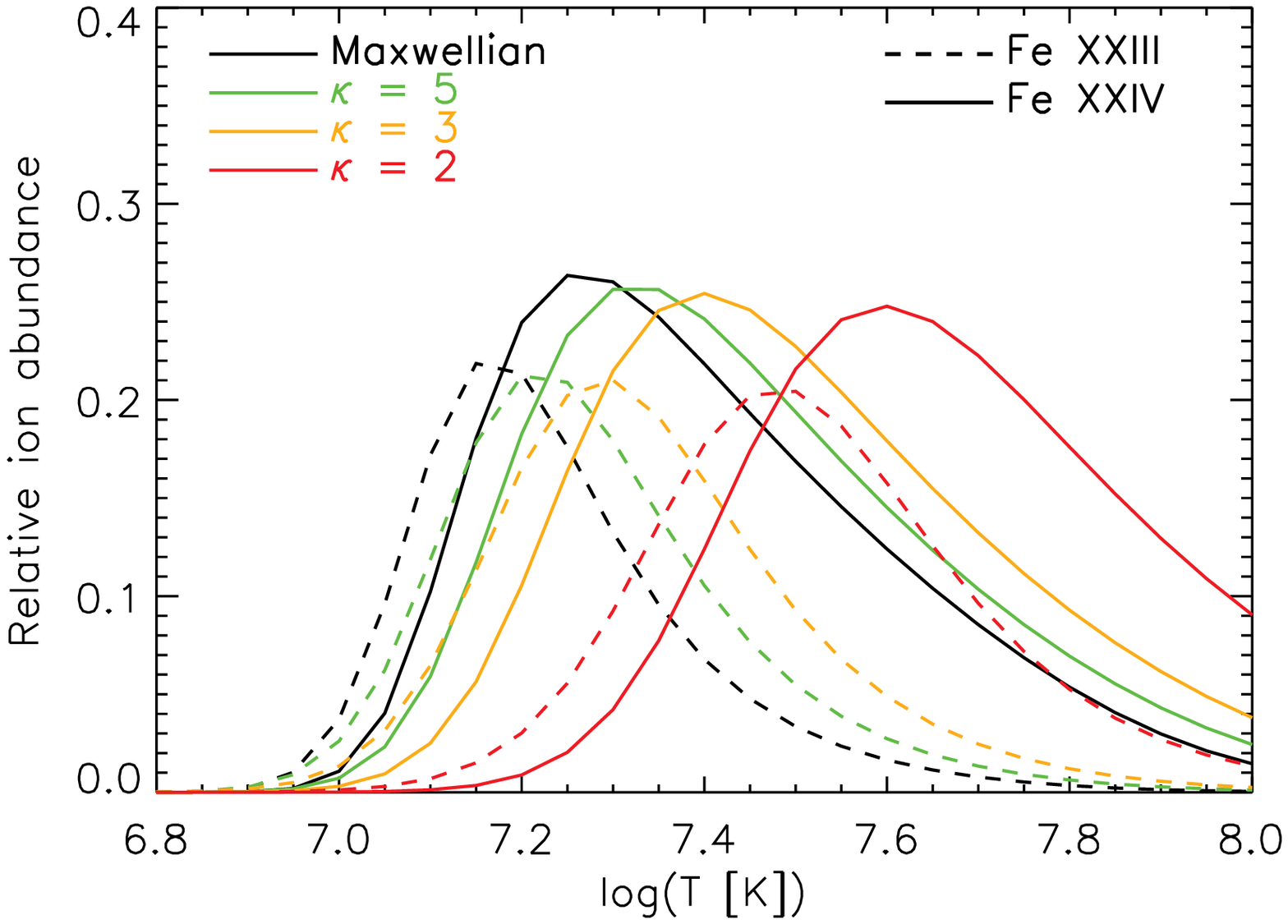}
\includegraphics[width=0.48\textwidth]{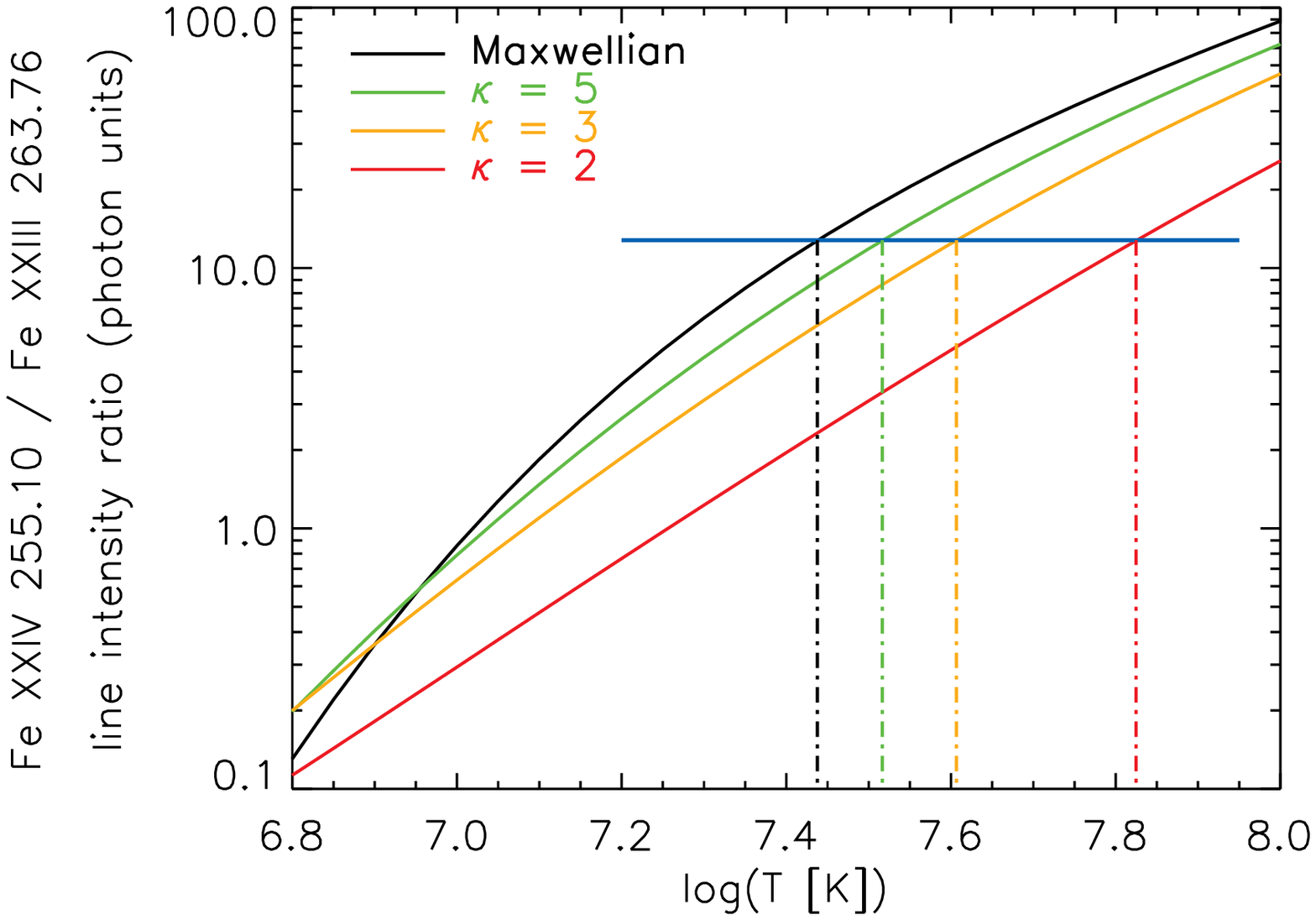}
\caption{\emph{Left panel}: fractional ion abundance for \fexxiii~and \fexxiv~for a Maxwellian distribution (black line) and non-Maxwellian distributions with different values of $\kappa$, as indicated in the legend. \emph{Right panel}: Theoretical \fexxiv~255.10/\fexxiii~263.76~\AA~ratio for electron Maxwellian (black line) and non-Maxwellian distributions with different values of $\kappa$, as indicated in the legend. The horizontal line indicates the observed ratio of 12.8 in box\,1.}
\label{Fig:temp_diagnostics}
\end{figure*}

\subsection{\fexxiv~192~\AA~line fitting}
\label{Sect:fe24_192}
The \fexxiv~192.03~\AA~line is the most intense of the observed EIS flare lines, as its centroid is close to the peak of sensitivity of the spectrometer. Unfortunately, the line was saturated in box\,1, where the \emph{RHESSI} HXR sources are located. Nevertheless, the line could be reliably observed in the region indicated by box\,2 in Fig.~\ref{Fig:overview}, which is also located along the plasma sheet feature. In contrast to the \fexxiv~255.10~\AA~line, both wings can be observed in the 192.03~\AA~spectrum. In most pixels of box\,2, the line shows the presence of a blend in the red wing and possibly a less intense one in the blue wing. The origin of both these blends is unclear. One may think that the blend on the red wing of the \fexxiv~192.03~\AA~line might be due to the \fexii~transition at 192.39~\AA. To investigate this possibility, we estimate the predicted intensity of the \fexii~192.39~\AA~line by measuring the intensity of another \fexii~transition at 186.88~\AA~and using the theoretical ratio of these two lines from CHIANTI. We find that the predicted intensity of \fexii~192.39~\AA~is significantly lower (by a factor of $\approx$50) than the intensity of the observed blend on the red side of the \fexxiv~192.03~\AA~line, and we can thus rule out this interpretation. Another possible explanation may be that we are observing an unknown transition or a red-shifted component of the \fexxiv~192.03~\AA~line. Such redshifted component might not be observed in the \fexxiv~255.10~\AA~spectra as in that case the line partially lies outside the spectral window. 

Figure~\ref{Fig:fit_192} shows an example of the \ion{Fe}{24} 192.03~\AA~spectrum at the raster exposure (or X-pixel) 12 and Y-pixel 225, which corresponds to Y-pixels~$\approx$~208--209 for the \fexxiv~255.10~\AA~observation, considering the offset of~$\approx$16.5 pixels between the SW and LW channels.
The four panels of Fig.~\ref{Fig:fit_192} show the results of different fitting procedures that we applied to the same spectrum (from left to right): a $\kappa$ fit with the one or two blends and a Gaussian fit with one or two blends. It should be noted that the blends in the $\kappa$ fit are assumed to be Gaussian, because their intensity lines is too low to determine the shape of their profiles, as well as to limit the number of free parameters of the fit. Figure~\ref{Fig:fit_192}  shows that a $\kappa$ distribution plus only one blend on the red wing of the line provides a very good fit of the observed spectrum, without the need of adding an extra component on the blue side of the \fexxiv~line. This is not the case for the Gaussian fit, where two extra components are required. These components also need to be very large to properly fit the line profile, resulting in a width for the \fexxiv~line which is narrower that the ones observed for the \fexxiv~255.10~\AA~and \fexxiii~263.76~\AA~lines, as shown in Fig.~\ref{Fig:fe_23_24}. While these two latter lines are too faint to perform a reliable fit of their wings (see the $\chi^2_\mathrm{red}$ value well below 1), their $FWHM_\textrm{G}$ values are estimated from the Gaussian fit to be around 0.30~\AA~for the \fexxiv~255.10~\AA~line and 0.27~\AA~for the lower temperature \fexxiii~263.76~\AA~line (corresponding to $w_\textrm{G}$ = 0.13~\AA~and 0.12~\AA~respectively), as indicated in Fig.~\ref{Fig:fe_23_24}. It is reasonable to assume that the width of the \fexxiv~192.03~\AA~would be at least as large as the width of the other \fexxiv~line. However, if we fix the minimum $FWHM_\textrm{G}$ of the \fexxiv~192.03~\AA~line to be at least equal to a conservative value of 0.26~\AA~(or equivalently $w_\textrm{G}$ = 0.11~\AA) in the 3 component Gaussian fit, this latter does not perform as well, with a value of $\chi^2_\mathrm{red}$ and residuals which are significantly larger than those obtained for the 3 component $\kappa$ fit (see Fig.~\ref{Fig:fit_192_width}). We consider this an additional indication that the \fexxiv~192.03~\AA~line, even if blended, is best fitted with a $\kappa$ rather than a Gaussian profile. 

The $\kappa$~values obtained from fitting the \fexxiv~192.03 \AA~and 255.10 \AA~lines are summarized in Fig.~\ref{Fig:kappa_map}, assuming two (i.e. one line blend, left panel) or three components (i.e. two line blends, right panel) for the fit of the 192.03~\AA~line.


%
%
%
\section{Temperature diagnostics based on the \fexxiv~255.10/\fexxiii~263.76~\AA~ratio}
\label{Sect:temp_diagnostics}
We use the ratio of the \fexxiv~255.10~\AA~and \fexxiii~263.76~\AA~lines to estimate the \textit{electron} temperature $T_\mathrm{e}$ of the emitting plasma in the location indicated by box\,1 in Fig.~\ref{Fig:overview} at around 15:59~UT. We convert the \fexxiv~and \fexxiii~intensities from DN to physical units (i.e. phot s$^{-1}$ cm$^{-2}$ arcsec$^{-1}$) using the radiometric calibration of \cite{DelZanna13}. As mentioned in Sect.~\ref{Sect:eis_data}, the EIS radiometric calibration may need to be revised to take into account of the instrument degradation after 2014. Nevertheless, problems in the calibration should not affect our results dramatically as we use lines which are close in wavelength and are included in the same CCD detector. We find that the ratio estimated using the radiometric calibration from \citet{DelZanna13} ($\approx$12.1) is very similar to the one obtained using the ground calibration ($\approx$12.8). This is not surprising as the correction factor shown in \cite{DelZanna13} appears to be similar in the wavelength interval around 250--260~\AA~\citep[see middle panel of Fig. 8 of ][]{DelZanna13}. We note that the in-flight calibration produces only a difference of $\approx$0.02 in the resulting log($T$\,[K]) estimated from the \fexxiv~255.1\,\AA~/ \fexxiii~263.8\,\AA~line intensity ratio. A ratio of 12.1 indicates a temperature of log($T$\,[K])~7.44 in the Maxwellian case (see the right panel of Fig.~\ref{Fig:temp_diagnostics}).

We also investigated the effect of taking into account non-Maxwellian conditions in the local plasma on the temperature diagnostic. Figure~\ref{Fig:temp_diagnostics} (left panel) shows the fractional ion abundance of the \fexxiii~and \fexxiv~lines for the Maxwellian and $\kappa$-distributions with different values of $\kappa$, as obtained using the KAPPA package \citep{Dzifcakova15} and CHIANTI v8. We note that these ionization equilibria were obtained by assuming \textit{electron} $\kappa$-distributions, which may not necessarily be the same as \textit{ion} distributions \citep[see Sect. 7.1 of][as well as our Sect. \ref{Sect:discussion}]{Dudik17a}. We note that the peak formation temperature of the \fexxiii~and \fexxiv~lines is strongly shifted to higher values for decreasing values of $\kappa$. In turn, the calculated \fexxiv~255.10/\fexxiii~263.76~\AA~ratios as a function of different $\kappa$, shown in the right panel of Fig.~\ref{Fig:temp_diagnostics}, are also shifted to higher $T_\mathrm{e}$ \citep[c.f.,][]{Dzifcakova18}. The horizontal blue line in this panel indicates the observed ratio in box\,1. Fig.~\ref{Fig:temp_diagnostics} shows that the temperature diagnostic varies significantly with $\kappa$, from log($T_\mathrm{e}$\,[K])\,$\approx$\, 7.44 for Maxwellian to 7.82 for $\kappa$\,=\,2, see dot-dashed lines in the right panel of Fig.~\ref{Fig:temp_diagnostics}. In particular, using the average value of $\kappa$ obtained from the fit of the \fexxiv~255.10~\AA~spectra in box\,1 ($\approx$2--3 see Sect.~\ref{Sect:fe24_255} and Fig.~\ref{Fig:kappa_map}), we estimate a temperature of the hot plasma at the plasma sheet feature of the X-class September 10 2017 flare to be between log($T_\mathrm{e}$\,[K])\,=\,7.6 and 7.8. This value is significantly higher than the corresponding Maxwellian one \citep[see also][Figures 5--6 therein]{Warren18}, by a factor of $\approx$2.4 for $\kappa$\,=\,2. These results are in line with the non-Maxwellian temperature diagnostics of \citet{Dzifcakova18}.

We also note that the diagnosed temperatures are above the corresponding ionization peaks of \fexxiv~independently of whether the conditions are Maxwellian or non-Maxwellian. Such high temperatures have been reported before \citep{Tanaka82} and in the context of our observation, they suggest that the the presence of cool line blends \citep[such as \ion{Fe}{17}, e.g. ][]{DelZanna08} on the wings of the \fexxiv~lines is unlikely, justifying the fitting procedures as described in Sect. \ref{Sect:fit}.

Using the temperature diagnostic obtained by the \fexxiv~and \fexxiii~line ratio in box 1, we can also provide a rough estimate of the electron number density $N_\textrm{e}$ in the plasma by using the emission measure ($EM_h$) obtained from the AIA images in the 193~\AA~filter:

\begin{equation}
EM_ h= \frac{I_\textrm{obs}}{R_\textrm{193\AA}(T)}
\label{eq:em}
\end{equation}

where $I_\textrm{obs}$ is the observed averaged intensity of the AIA 193~\AA~image in box\,1 at 15:59~UT, expressed in DN s$^{-1}$. $R_\textrm{193\AA}(T)$ is the AIA response function in the 193~\AA~filter, which can be obtained by using the effective area provided by the SSW routine \texttt{aia\_get\_response.pro}, atomic data from CHIANTI v8 \citep{Dere97, DelZanna15} and photospheric abundances from \cite{Asplund09}, following the method described in the appendix of \cite{DelZanna11}. Assuming the range of temperatures (log($T$\,[K])\,$\approx$\,7.3--7.7) obtained from the \fexxiv~255.10/\fexxiii~263.76~\AA~ratio for Maxwellian and $\kappa$ distributions, we obtain $EM_h$ values varying between 5.2 $\times$ 10$^{31}$ cm$^{-5}$ and 1.8 $\times$ 10$^{32}$ cm$^{-5}$.

The plasma density can thus be estimated by using the following formula: 
\begin{equation}
N_\textrm{e} \approx \sqrt{\frac{EM_\mathrm{h}}{N_\mathrm{H} \times N_\mathrm{e} \times h}}
\label{eq:density}
\end{equation}

where \emph{h}~is depth of the plasma sheet, estimated by \cite{Warren18} to be $\approx$10$^{8.9}$ cm. The hydrogen density \emph{N}$_\textrm{H}$ can be expressed as 0.83 \emph{N}$_\textrm{e}$ in a fully ionized gas with helium abundance relative to hydrogen A(He) = 0.1. Using Eq.~\ref{eq:density} and the values of $EM_h$ above, we obtain densities of the order of 2.8--5.3~10$^{11}$ cm$^{-3}$ in the plasma sheet (at the location indicated by box 1 in Fig.~\ref{Fig:overview}) during the impulsive phase of the September 10 2017 flare. These densities can be directly compared with the values obtained from the analysis of the \emph{RHESSI} spectra, as will be described in Sect.~\ref{Sect:rhessi}.

\section{Analysis of \emph{RHESSI} data}
\label{Sect:rhessi}
\emph{RHESSI} \citep{Lin02} started detecting the flare at $\approx$~15:53 UT, while the very beginning of the flare
was missed by the instrument, which was in the night part of its orbit. During the impulsive phase until $\approx$~16:00~UT,
the lightcurves of the HXR
emission show several bursts up to $\sim$~300~keV, whereas the emission at energies below $\sim$~25~keV rises smoothly.
We analyzed \emph{RHESSI} spectra and images observed around 15:59 UT, when the EIS slit passed over the \emph{RHESSI} HXR source.
Despite the thick attenuator, i.e. the A3 state, was in place at that time,
the \emph{RHESSI} data are still affected by pile-up effects which distort the
spectra formed in the $\approx 30-50$~keV range. Currently, the pile-up cannot be corrected for when
reconstructing the \emph{RHESSI} images. Therefore, in Fig.~\ref{Fig:overview}
the 30--50~keV energy range is omitted and the \emph{RHESSI} sources are shown at energies below and above it. 
Figure~\ref{Fig:overview} shows that the \emph{RHESSI} sources are co-spatial with the upper parts of the bright loops as seen e.g. in the AIA~193~\AA~filter.

The \emph{RHESSI} spectra were analyzed individually for for detectors  3 and 8 (which seem to be less affected by pile-up than detectors 1 and 6)
and fitted within the $12-250$~keV energy range.
Energies below 12~keV were not included in the spectral analysis due to unknown instrumental effect modifying spectra
in the A3 state. In order to take into account the pile-up effects for the spectral analysis,
\verb(pileup_mod( fitting function was used. We analyzed the \emph{RHESSI} spectrum accumulated during the time interval
15:59:04-15:59:16 UT, corresponding to the EIS box~1, in Fig.~\ref{Fig:overview}.

\begin{figure}
  \centering
\includegraphics[width=0.45\textwidth,bb=30 180 540 530,clip]{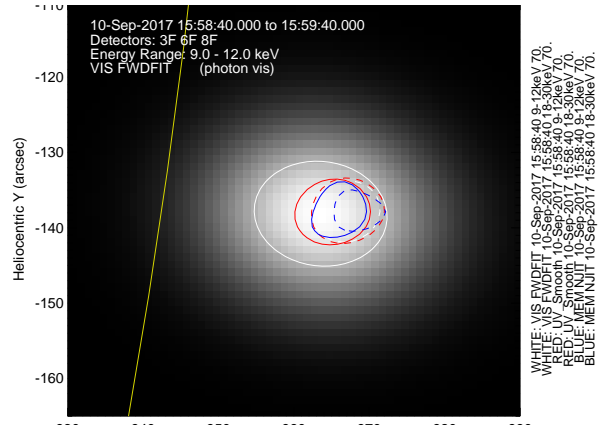}
  \caption{\emph{RHESSI} sources in the $9-12$~keV (full lines) and $18-30$~keV (dashed lines) range
    reconstructed by three image algorithms: VIS FWDFIT (white), UV\_Smooth (red), and MEM NJIT
    (blue). The background
  image is VIS FWDFIT image in the $9-12$~keV range. Countours are displayed at the 70$\%$ intensity level.}
\label{Fig:rhessi_thermal}
\end{figure}

\begin{figure}
  \includegraphics[width=0.45\textwidth]{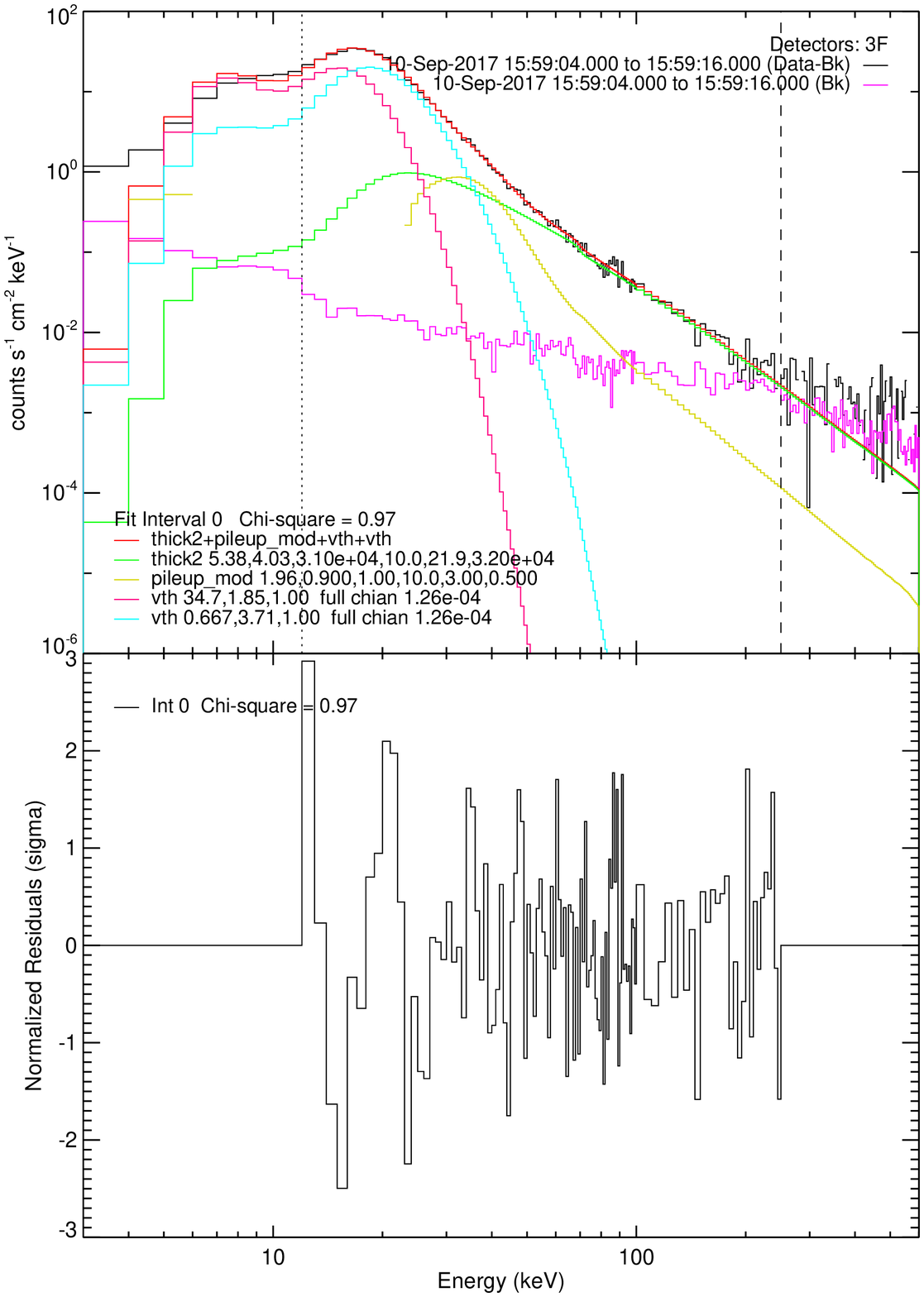}
  \includegraphics[width=0.45\textwidth]{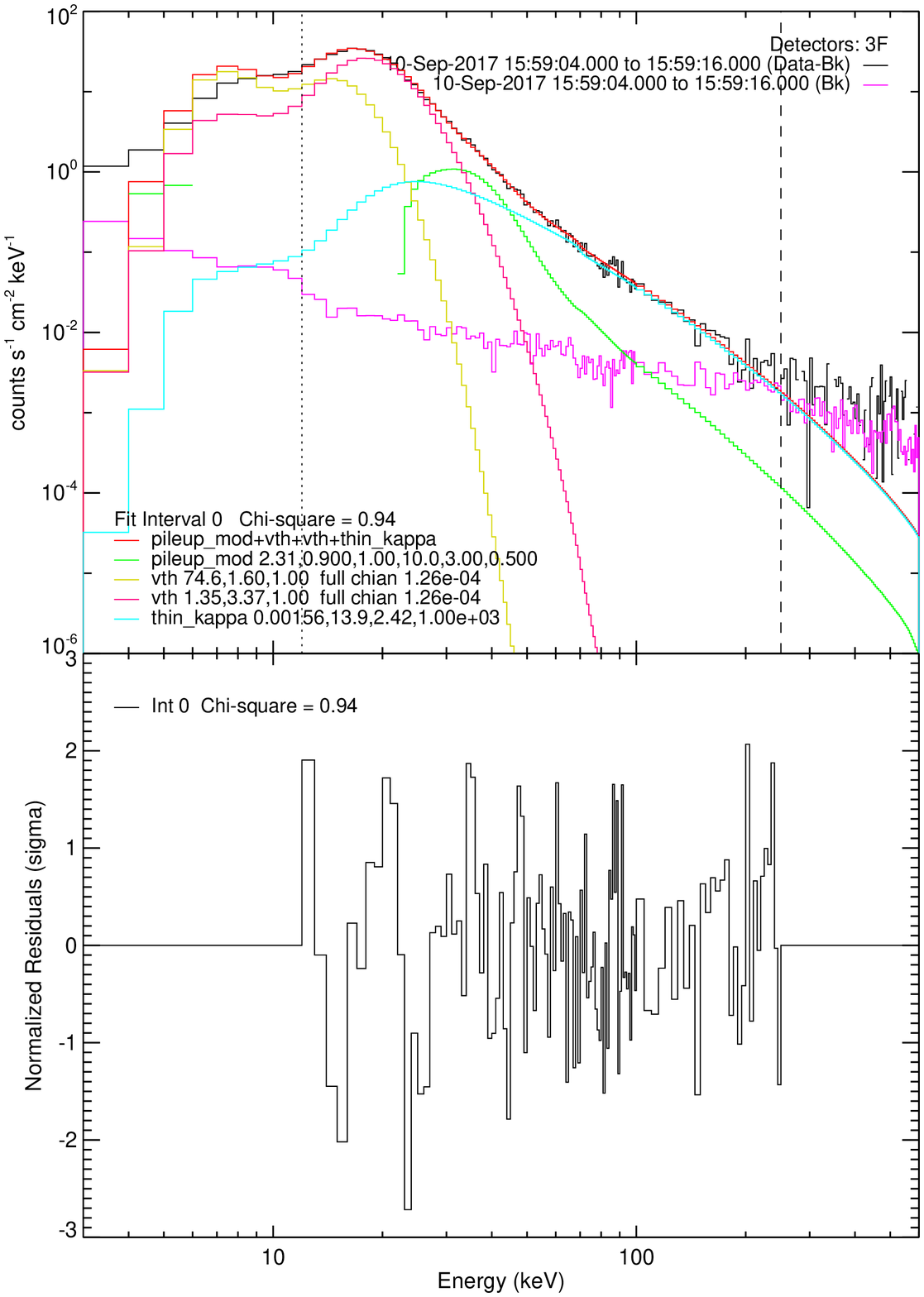}
  \caption{\emph{RHESSI} spectra of 3F detector: Left: Forward-fitted spectrum using two thermal components, vth,
    a correction for pile-up effects, pileup\_mod, and a thick-target power-law beam, thick2. Residua of the modelled and
    observed spectra are showed below the spectrum. Right: The same as in the left panel but using 
    a kappa distribution in the thin-target approximation, thin\_kappa, instead of the thick-target beam component.}
  \label{Fig:rhessi_spectra}
\end{figure}

 The fits have revealed that a multi- or a double-thermal component dominating emission up to $\sim 50$~keV
  is needed to describe the \emph{RHESSI} spectrum well. Both models are similar in terms of $\chi^{2}$ and residuals.
  Fitting with a multi-thermal component, assuming that differential emission measure
  has a power-law dependence on temperature ($\sim T^{-\alpha}$), results in: a power-law index $\alpha$ $\approx$~6.3, differential emission
  measure at $T=23$~MK (2~keV) $DEM_{23} = 15\times 10^{49}\,\mbox{cm}^{-3}\,\mbox{kev}^{-1}$,
  and a maximum temperature $T_{\mbox{m}}=55-92$~MK. On the other hand,
  the fit using a double-thermal component reveals the presence of a super-hot component, $T_{\mbox{sh}}=43-56$~MK,
which dominates the HXR spectra in $\sim 18-30$~keV energy range, see Fig.~\ref{Fig:rhessi_spectra}.
The \emph{RHESSI} source reconstructed at this energy range
is shifted towards higher coronal heights compared to the thermal source, $T=21$~MK at lower energies $9-12$~keV (see Fig.~\ref{Fig:rhessi_thermal}), similarly as the super-hot sources reported by \citet{ca10,ca15}. The separation between thermal and super-hot sources is rather small, around 1-2\arcsec, but is seen consistently in all three different algorithms used for the image
reconstruction (MEM NJIT, UV\_Smooth, and VIS FWDFIT). Although source sizes and positions slightly differ between
algorithms, the reconstructed \emph{RHESSI} sources are similar in terms of modeled and observed modulation
profiles and visibilities. 
We estimated a source area $S$ and volume $V$ from the $50\%$ intensity contour of
the reconstructed images, taking $V=4/3\pi(S/\pi)^{3/2}$. Then, using the fitted emission measure of
the thermal and super-hot components,
$EM=(35-46)\times 10^{49}\,\mbox{cm}^{-3}$ and $EM_{\mbox{sh}}=(0.28-0.76)\times 10^{49}\,\mbox{cm}^{-3}$,
and their volume, the electrons density can be determined. The density of the thermal source is
high, $N_\mathrm{e}=(4.3-13)\times 10^{11}\,\mbox{cm}^{-3}$, whereas the density of the super-hot source is
$N_{\mbox{e,sh}}=(0.86-2.2)\times 10^{11}\,\mbox{cm}^{-3}$, again similarly as the one reported by \citet{ca10} for the 2002 July 23 X4.8 flare.
These densities are consistent with the values ($\approx 2.8-5.3\, 10^{11}\,\mbox{cm}^{-3}$)
obtained in Sect.\ref{Sect:temp_diagnostics}, using the estimated emission measure in the AIA 193~\AA~channel
and temperature diagnostics from the EIS line ratio. 
Such a high density source can be viewed as an example of a coronal
thick-target source type discovered by \citet{ve04}. Assuming source half-length $L =\sqrt{S/\pi}$
and the relation for the beam stopping energy $E_{\mbox{stop}}\approx\sqrt{10^{-17} n L}$ \citep[e.g.][]{kr08},
the thermal source region is capable of collisional stopping of electrons of energies up to about $57-72$~keV.
Indeed, the \emph{RHESSI} loop-top source in the $50-100$~keV range overlaps with the \emph{RHESSI} thermal source and has a comparable size. However, we note that a \emph{RHESSI} footpoint source in energies above ~30 keV does exist \cite{Gary18}. This could suggest that the footpoint source might not not in the same
loop/magnetic field as the loop-top source.

Given the nature of \emph{RHESSI} sources, a thick-target component at energies
above $\sim 50$~keV was used in spectral analysis of spatially integrated \emph{RHESSI} spectra.
The observed \emph{RHESSI} spectrum can be fitted with an injected electron power-law beam of $\delta$ $\approx$~4,
which is significantly harder than the values reported by \citet{ve04}.
Although the \emph{RHESSI} source in the $100-300$~keV range overlaps with the thermal source, it has a more elongated shape
and cannot be interpreted as a thick-target region for electrons of such energies. Using the half-length measured
across the $50\%$ intensity contour, the stopping energy is in the $77-130$~keV range. Therefore, it is also relevant
to use the thin-target approximation. Figure~\ref{Fig:rhessi_spectra} shows such an example for a thin-target kappa
distribution \citep[see][]{kasparova2009,Oka13}.
Further, the spectra can be fitted with a thin-target power-law beam equally well in terms of $\chi^2$
and residuals.  While the power-law index thus obtained is $\approx$~2.5, $\kappa$ ranges between 2.1 and 2.4.
These values are consistent to the ones derived for ions from the EIS line fitting (Sect.\ref{Sect:fit}).

Note that at energies below $50$~keV the \emph{RHESSI} spectrum can be fitted by a single kappa distribution instead of two
thermal components. Although such a component has fewer free parameters, due to the different source sizes
and positions in the energy ranges $9-12$~keV and $18-30$~keV where each of the thermal components dominates,
we excluded such a fitting function as inappropriate. Finally, the \emph{RHESSI} spectrum cannot be
  fitted well enough with two components only, e.g. with a thermal component
  and a beam single power-law/$\kappa$ component \citep[as in][]{Oka15}. A reasonable fit
  can also be obtained for a double power-law beam with a steep (soft) spectrum above $\sim$70~keV only, i.e.
  mimicking the nature of a multi- or double-thermal component. Such a fit was not interpreted
  as more appropriate.

\section{Discussion and Interpretation}
\label{Sect:discussion} 

We now proceed to discuss the possible interpretation of the observed line profiles and \emph{RHESSI} spectra. 

\citet{Jeffrey16} listed three possible interpretations to the observed $\kappa$ emission line profiles in solar flares:
\begin{enumerate}
\item Plasma is multithermal along the line of sight;
 \item Fe ions are accelerated isotropically to a non-Maxwellian distribution of velocities;
 \item Turbulence is present, in terms of macroscopic parameters ($T$, $n_\mathrm{e}$, or velocities).
 
\end{enumerate}
We shall deal with these in the following subsections.

\subsection{Multithermal plasmas}
\label{Sect:discussion_multithermal}

The interpretation (1) is based on the fact that any non-Maxwellian distribution with a thermal core and a tail can be represented by a linear combination of Maxwellians with different temperatures. For a $\kappa$-distribution, this Maxwellian decomposition and its coefficients was provided by \citet{Hahn15}. In terms of the line profile fitting, one should fit a series of Gaussians to the line profile, each representing a Maxwellian at a given temperature. We have seen in Sect. \ref{Sect:fe24_255} that such fitting is unconstrained, with the exception shown in the right panel of Figure \ref{Fig:fit_255_exp8}. Using the widths of such double-Gaussian fit, $w_\mathrm{G1,2}$, we obtain equivalent ion temperatures $T_\mathrm{G1,2}$ (also called Doppler temperatures) by using the formula \citep[compare Eqs. (8) and (10) of][]{Dudik17a}:
\begin{equation}
 w_\mathrm{G}^2	-w_\mathrm{instr}^2 = \frac{k_\mathrm{B} T_\mathrm{G} }{m_\mathrm{Fe}} \frac{\lambda_0^2}{c^2}\,,
 \label{Eq:T_G}
\end{equation}
where we subtracted the $w_\mathrm{instr}^2$ factor to account for the instrumental broadening. Eq. (\ref{Eq:T_G}) yields 
yields $T_\mathrm{G1,2}$\,=\,62 and 99\,MK for the two Gaussians, respectively. Both are significantly higher (by factors of $\approx$2.3 and 3.6) than the corresponding $T_\mathrm{e}$ derived from the \ion{Fe}{24}\,/\,\ion{Fe}{23} line intensity ratio (Sect. \ref{Sect:temp_diagnostics}). If the plasma were really multithermal with dominant temperature components at $T_\mathrm{G1}$ and $T_\mathrm{G2}$ (assuming equilibrium between electrons and ions), the $T_\mathrm{e}$ obtained should be correspondingly higher. Since it is not, we suggest that the multithermal interpretation is not likely.

\subsection{Accelerated ions}
\label{Sect:discussion_acc_ions}

The interpretation (2), namely that ions are accelerated, is in line with recent theoretical modeling of \citet{Li17}, who showed that ions can be accelerated preferentially to electrons in flare conditions. That the ion profiles could be due to accelerated ions has also been thoroughly examined by \citet{Jeffrey17}, who used the formulae derived by and adapted from \citet{Sigmar71} and \citet{Stix72} to calculate the thermalization time for accelerated ions; in their case, \ion{Fe}{16} and \ion{Fe}{23}. The thermalization timescale $\tau_f$ was derived to be $\approx$10$^{-2}$\,s.
In our case, using the values of $N_\mathrm{e}$\,=\,$5 \times 10^{11}$\,cm$^{-3}$ and $T_\mathrm{e}$ derived in Sect.~\ref{Sect:temp_diagnostics}, the corresponding ion thermalization timescales would be even shorter by about 2 orders of magnitude. 

However, we note that the formula (13) of \citet{Jeffrey17} is applicable only for ion velocities $v$ satisfying $v_\mathrm{th,i} \ll v \ll v_\mathrm{th,e}$, where $v_\mathrm{th}$ are ion and electron thermal speeds, equal to $\sqrt{2k_\mathrm{B}T_\mathrm{e,i}/m_\mathrm{e,i}}$, respectively. In our case, the condition $v_\mathrm{th,i} \ll v$ is broken: If we suppose that the ions are accelerated to a $\kappa$-distribution corresponding to $\kappa$\,$\approx$\,2 profiles shown in Figures \ref{Fig:fit_255_exp9}, we obtain, as an order-of-magnitude estimate, Doppler velocities $v$\,$\approx$\,300 km\,s$^{-1}$ for the far wings of the \ion{Fe}{24} 255.1\,\AA~line at $\Delta\lambda$\,$\approx$\,0.25\,\AA, since $\Delta\lambda/\lambda_0 = v/c$.
If we assume that the ions have the same temperature as electrons, i.e., log($T_\mathrm{e}$ [K])\,=\,7.4--7.8 derived in Sect. \ref{Sect:temp_diagnostics}, we obtain ion thermal velocities of 90--140\,km\,s$^{-1}$; i.e., of the same order-of-magnitude. Thus, the formula (13) of \citet{Jeffrey17} for calculation of ion thermalization timescales may not be entirely applicable in our case.

\citet{Jeffrey17} invoked the results of \citet{Bian14} to conclude that if the line profiles are due to accelerated ions, then the ions must be accelerated locally and continously during the flare. This conclusion is based on the fact that, according to \citet{Bian14}, the $\kappa^*$ index \citep[with $\kappa^*$\,=\,$\kappa+1$,][see Sects. 3 and 4.1 therein]{Livadiotis09} is a parameter describing the ratio of the competing acceleration and collisional timescales, $\kappa^* = \tau_\mathrm{acc}/2\tau_\mathrm{coll}$. Such conclusion is natural, since the short ion thermalization timescales mean that if the acceleration process is turned off, the ions thermalize much faster compared to EIS exposure times; therefore, if non-Gaussian lines are observed, the acceleration process must persist.

If our line profiles are also due to accelerated ions, then the conclusion of \citet{Jeffrey17} is again valid and straightened both in terms of acceleration being local as well as continuous. This is because in our case, the ion thermalization timescales are even shorter, while we get similar values of $\kappa^*$. Since the thermalization timescale is proportional to $\tau_\mathrm{coll}$, and $\tau_\mathrm{coll} = \tau_\mathrm{acc}/2\kappa^*$, and since the thermalization length $L_f$\,=\,$v \tau_f$ is proportional to the product of ion velocity and thermalization timescale, our smaller thermalization timescale \citep[compared to][]{Jeffrey17} means both shorter acceleration timescales and shorter thermalization lengths.

We also emphasize that similar $\kappa$ values were found for the ion (from the EIS \fexxiv~line profiles) as well as electron (from \emph{RHESSI}) distributions. This consistency suggests that the ions were accelerated and that the acceleration process produced both fast electrons and ions with similar distribution functions. 

To summarize, the interpretation that ions are accelerated locally and continuously during the flare, remains a valid candidate for our observation.

\subsection{Turbulence}
\label{Sect:discussion_turbulence}

Macroscopic turbulence is often invoked as an explanation for the non-thermal line widths, i.e., widths larger than the corresponding thermal ones given by Eq. (\ref{Eq:T_G}). The reasons for the invocation of turbulence in flare observations are well described e.g. in Sect. 3 of the review of \citet{Antonucci89}. One typical example is given by the observation of large non-thermal widths 
in the spectra of blueshifted high temperature lines during chromospheric evaporation, which are often explained in terms of turbulent mass motions, or alternatively superposition of unresolved flows 
 \citep[e.g.][]{Milligan11,Polito15}.

Rather than chromospheric evaporation, the non-thermal broadening could also be related to HXR emission during flares, as the onset of non-thermal broadening occurs simultaneously with detection of low-intensity HXR emission, and can even precede the strong HXR bursts \citep[e.g.]{Antonucci89}. Alternatively, the non-thermal
broadening could be associated to redshifts (due to small bulk downflows),
as reported by \citet{Jeffrey17},  which could drive
turbulence. In our case, it is difficult to establish the presence of true donwflows along the flaring structures due to the off-limb geometry, and the uncertainty of the EIS wavelength calibration \citep{Kamio10}.

The observed consistency of non-thermal broadenings observed in lines of ions of different elements and  \citep[e.g.][Table 1 therein]{Doschek79}, and the absence of variation with the flare position on disk or limb (in contrast to what expected by assuming that the broadening is caused by superposition of flows) was also used as an argument in favor of turbulent (random) motions. Further, in these early studies, large differences between equivalent ion temperatures and electron temperatures derived from line ratios were observed for several minutes \citep[e.g.][]{Antonucci89}, suggesting that the broadening could not be caused by an effective ion-electron temperature difference, as this could not be maintained for such a long timescale.

But is this line of reasoning valid in our case? To test this, we first estimate the equivalent ion temperatures for the ion $\kappa$-distributions corresponding to the observed line profiles. We take the value of $w_\kappa$\,=\,0.15\,\AA~for the \fexxiv~line at 255.1\,\AA~from Fig. \ref{Fig:fit_255_exp9} and calculate the equivalent $T_{\kappa}$ using the formula (compare Eqs. (6) and (11)) of \cite{Dudik17a}:
\begin{equation}
 w_\mathrm{\kappa}^2	= \frac{k_\mathrm{B} T_{\kappa} }{m_\mathrm{Fe}} \frac{\lambda_0^2}{c^2}\,.
 \label{Eq:T_kappa}
\end{equation}
Note that the $w_\kappa$ and $w_\mathrm{G}$ were derived to have the same physical meaning as if interpreted in terms of the equivalent ion temperature. This stems from the temperature having the same physical meaning for a Maxwellian and a $\kappa$-distribution. We also note that, unlike for $w_\mathrm{G}$ (Eq.\,\ref{Eq:T_G}), the instrumental width is already accounted for since the fit is a convolution (see Eq.\,\ref{Eq:convolution}). Using the above formula, we obtain $T_\kappa$\,=\,205\,$\pm$22\,MK, or log($T_\kappa$\,[K])\,$\approx$\,8.3. Similarly, for the lowest value of $w_\kappa$\,=\,0.10 (Fig. \ref{Fig:fit_255_exp8}), we would still obtain $T_\kappa$\,=\,99\,$\pm$\,3\,MK. The uncertainties in these temperatures are calculated using the corresponding uncertainties of $w_\kappa$.

An estimate of the upper limit on the uncertainty of  $T_\mathrm{e}$ derived from the \fexxiv~/\fexxiii~line intensity ratio can be obtained by assuming that both the \fexxiv~255.1\,\AA~and \fexxiii~263.8\,\AA~line intensities contain not only the photon noise uncertainty, but also the $\approx$20\% calibration uncertainty \citep{Culhane07} as well, despite these lines being both observed in the long-wavelength channel of EIS. Doing so would yield electron temperatures of 27\,$\pm$4\,MK assuming Maxwellian, 39\,$\pm$7\,MK for $\kappa$\,=\,3, and 66\,$\pm$10\,MK for $\kappa$\,=\,2, respectively. We emphasize that in general assuming that the ions are formed under non-Maxwellian conditions results in a higher formation temperature (and thus broader thermal widths) than expected in Maxwellian conditions. In turn, this decreases the amount of non-thermal width that needs to be invoked to explain the observed profiles.
Nevertheless, the temperatures listed above are still too low compared to  the equivalent ion temperatures which are required to explain the observed large widths in this work. The conclusion is thus that turbulence is still a possible explanation for the observed line widths $w_\kappa$ as well as the non-Gaussian $\kappa$ profiles. We note that if the turbulent diffusion coefficient is inversely proportional to velocity, a $\kappa$-distribution is formed \citep{Bian14}.

Thus, we estimate the ``non-thermal'' widths $w_\mathrm{nth}$ for the non-Gaussian profiles. To a first approximation, this can be done by setting \citep[c.f.,][]{Dudik17a}
\begin{equation}
 w_\mathrm{nth}^2 = w_\kappa^2 -w_\mathrm{th}^2\,.
 \label{Eq:w_nth}
\end{equation}
This equation is not exact; rather, it gives a lower limit of the $w_\mathrm{nth}$.  This is since a convolution of two $\kappa$-distributions with the same $\kappa$ but different $w_\mathrm{th}$ and $w_\mathrm{nth}$ is \textit{not} a $\kappa$-distribution of the same $\kappa$. This arises from the fact that a sum is present in the $(1 +(\lambda-\lambda_0)^2/2\kappa w^2)^{-\kappa}$ sub-integral functions. However, the resulting convolution is slightly wider, but not too different from those obtained using the approximation (\ref{Eq:w_nth}).

For the fit results from Fig.\,\ref{Fig:fit_255_exp8}, i.e., $\kappa$\,$\approx$\,3, $w_\kappa$\,=\,0.10, we obtain for $T$\,=\,39\,MK that $w_\mathrm{th}$\,=\,0.065\,\AA, from which subsequently $w_\mathrm{nth}$\,=\,0.076\,\AA~(or $\approx$126\,\kps), a factor of almost $\approx$1.2 larger than $w_\mathrm{th}$. Similarly, for the fit results of Fig.\,\ref{Fig:fit_255_exp9}, i.e., $\kappa$\,$\approx$\,2, $w_\kappa$\,=\,0.15, we obtain for $T$\,=\,66\,MK that $w_\mathrm{th}$\,=\,0.085\,\AA, from which subsequently $w_\mathrm{nth}$\,=\,0.124\,\AA~(or $\approx$205 km\,s$^{-1}$), again a factor of almost $\approx$1.5 larger than the corresponding $w_\mathrm{th}$. 

We note that the non-thermal velocity of 205\,km\,s$^{-1}$ is among the largest reported in flares, being even higher than the value of $\approx$160\,km\,s$^{-1}$ reported from X-ray and EUV spectra \citep[e.g.,][]{Doschek79,Doschek80X,Feldman80M,Antonucci82,Antonucci89,Landi03,DelZanna08}. To our knowledge, there are not many reports of higher non-thermal velocities. For example, \citet[][Figure 3 therein]{Tanaka82} reported such velocities from \ion{Fe}{26} spectra (but not \ion{Fe}{25}), decreasing from 250\,km\,s$^{-1}$ during the course of their flare. \citet{Antonucci86} reported velocities of 220\,\kps. At present, similar non-thermal line widths were seen with EIS in the \fexxiii~line previously by \citet{Lee17}, see their Fig. 9.

\subsection{Super-hot ions due to collisions \\ with high-energy electrons?}
\label{Sect:discussion4}

Finally, we turn our attention to a possible explanation not considered by \citet{Jeffrey16,Jeffrey17}, which is that the $\kappa$ distributions might be caused by collisions of super-hot ions with high-energy electrons. This scenario is suggested by the fact that the high-energy tail in the \emph{RHESSI} spectra could be fitted with a $\kappa$-distribution with $\kappa$\,values of \,2.1--2.4 (Sect. \ref{Sect:rhessi}), which are similar to the ones obtained from some line profiles (see, e.g., Fig. \ref{Fig:fit_255_exp9} and \ref{Fig:kappa_map}). 
Although the fitting of the high-energy tail observed by \emph{RHESSI} yields a well-constrained $\kappa$, the corresponding temperatures are not well-constrained. This is due to the fact that the quasi-thermal core of such $\kappa$-distribution occurs at low energies, of about $\approx$20--30\,keV (cyan line in the top right panel of Fig. \ref{Fig:rhessi_spectra}), where the spectrum is dominated by the thermal components and pileup. We determined that electron temperatures of about 20--160\,MK are still compatible with the observed high-energy tail, without an appreciable change in the $\chi^2$ and residuals of the \emph{RHESSI} spectrum fit. In fact, the top right panel of Fig. \ref{Fig:rhessi_spectra} shows a fit using a temperature of 14\,keV, equivalent to about 160\,MK. These temperatures are comparable, at least to within an order of magnitude, to the equivalent ion temperatures $T_\kappa$ derived in Sect. \ref{Sect:discussion_turbulence}.

Could such temperatures be realistic? In the 2D particle-in-cell simulation of \citet{Karlicky11}, electron temperatures of 60--120\,MK are reached during merging of plasmoids occurring due to tearing instability in the current sheet. In addition, the velocity distribution could show high-energy tails, although details depend on the location \citep[see Fig. 6 in][]{Karlicky11}.
Such process could possibly occur within our flare, and we note that the location of the observed non-Gaussian lines (Figure \ref{Fig:kappa_map}), i.e., the top of the flare arcade as well as in the plasma sheet feature, would both be appropriate with respect to the geometry of \citet{Karlicky11}, as plasmoids can also impact the top of the flare arcade and merge with it \citep[see also, e.g.,][]{Jelinek17}.

For the above reasons, as well as the fact that the \emph{RHESSI} thermal coronal source is thick-target for energies $\lesssim$57--72\,keV (Sect. \ref{Sect:rhessi}), we proceed to speculate whether the high equivalent ion temperatures could be due to impact of these high-energy $\kappa$\,=\,2 electrons. We first calculate the electron thermalization time due to collisions with both electrons and ions, using Eq. (3.50) of \citet{Goedbloed04}
\begin{equation}
 \tau_\mathrm{e} \approx \frac{1.09 \times10^{10}}{\mathrm{ln} \Lambda} \frac{\widetilde{T_\mathrm{e}}^{3/2}}{Z_f N_\mathrm{e}}\,,
 \label{Eq:tau_e}
\end{equation}
with $T_\textrm{e}$ = 3 keV ($\approx$~35 MK) and $N_\textrm{e}$ = 5 10$^{11}$ cm$^{-3}$.
where ln$\Lambda$\,$\approx$\,10 is the Coulomb logarithm, $\widetilde{T_\mathrm{e}}$ is electron temperature in keV, $N_\mathrm{e}$ is electron number density in cm$^{-3}$, and $Z_f$ is the charge of ions involved, which we take $Z_f=1$. Note that the $\tau_\mathrm{e}$ scales with $\widetilde{T_\mathrm{e}}^{3/2}$ due to the fact that progressively higher-energy electrons are less collisional. Considering now that the background plasma impacted by the beam has $\widetilde{T_\mathrm{e}}$\,=\,3\,keV ($T_\mathrm{e}$\,=\, and $N_\mathrm{e}$\,$\approx$\,5$\times$10$^{11}$ cm$^{-3}$, we obtain $\tau_\mathrm{e}$\,=\,8\,$\times$10$^{-3}$\,s. The corresponding electron-ion temperature equilibration time, i.e. the time at which electrons and ions reach thermal equilibrium with the same temperature, is longer by a factor $m_\mathrm{i}/2m_\mathrm{e}$ \citep[see Eq. (3.55) of][]{Goedbloed04}. Considering the Fe ions, we obtain that the temperature equilibration time is $\approx$388\,s.

Therefore, in principle, if the electron beam persists for such timescales, the ions could possibly be heated by impacting electrons to temperatures of the electron high-energy tail. However, we remind the reader that the temperature of the $\kappa$\,$\approx$\,2 high-energy tail is not well-constrained from \emph{RHESSI}.

Finally, we estimate the amount of energy that could be transferred to the ions from the impacting high-energy electron tail. To do so, we calculate the amount $\delta E_\mathrm{e}$ of the energy contained in the tail. This can be done by integrating the corresponding \emph{RHESSI} spectral component, resulting in $\delta E_\mathrm{e}/\delta t$\,=\,3.7\,$\times$10$^{28}$\,erg\,s$^{-1}$. The corresponding temperature gain by particular ions can then be estimated as
\begin{equation}
 \Delta T_\mathrm{i} \sim \frac{\delta E_\mathrm{e}}{\delta t} \frac{\Delta t A_\mathrm{Fe} }{N_\mathrm{e} V k_\mathrm{B}}\,,
 \label{Eq:T_gain}
\end{equation}
where $\Delta t$ is the duration of the high-energy tail, $A_\mathrm{Fe}$ is the iron abundance, and $V$ is the ambient volume, where the energy exchange occurs. Using the photospheric value of $A_\mathrm{Fe}$\,$\approx$\,3.2\,$\times$\,10$^{-5}$ \citep{Asplund09}, estimating $V$ from the size of the \emph{RHESSI} thermal source at 9--12\,keV as $V$\,$\approx$\,2.7\,$\times$10$^{26}$\,cm$^3$, and using $\Delta t$\,=\,11\,min as an upper-limit estimate from the \emph{Fermi} lightcurves \citep[see also][]{Long18}, we obtain $\Delta T_\mathrm{i}$\,=\,0.05\,MK. 
This value is too low for ambient ions to be efficiently heated by the high-energy electron tail. The value could be increased only by considering that the ambient volume $V$ is much smaller; however, decreasing it by even two orders of magnitude would still produce inefficient heating. We therefore conclude that this mechanism is not a likely explanation of the observed ion emission line profiles.

\section{Summary}
\label{Sect:Summary}

We reported on the \emph{Hinode}/EIS observations of wide and non-Gaussian profiles of \fexxiv~EUV lines during the impulsive phase of the 2017 September 10 X8.3--class solar flare. These lines are the hottest observed by EIS. We speculate that such a strong departure from Gaussian profiles in the high temperature lines could be observed in this large flare event (and not in previous X-class flares observed by EIS) thanks to the combination of: (1) very intense line profiles, with wings that could be properly fitted; (2) the ideal  location of the flare on the limb, which allowed us to observe the high temperature emission in the plasma sheet without contamination from the loop emission; (3) the favorable position of the EIS slit above the flare loops. 

The profiles of the \fexxiv~lines could be reliably fitted with a $\kappa$-distribution with low values of $\kappa$, in the range of $\approx$1.7--3.3. Different $\kappa$-values provide information about the number of particles in the high-energy tail \cite[see e.g.][Fig.1]{Oka13}.For instance, an electron distribution with $\kappa$=2 means that $\approx$~35\% of the particles are accelerated and they carry $>$~80\% of the energy. The non-Gaussian line profiles were found in the location of the top of the flare arcade as observed by \emph{SDO}/AIA in boxes 1 and 2 of Fig.~\ref{Fig:overview}, at the bottom and along the plasma sheet feature at about 15:59\,UT respectively.  The location in box 1 is coincident with the maximum intensity of the \emph{RHESSI} thermal and non-thermal sources at energies of 6 to 300\,keV, and is consistent with the location of the EOVSA microwave gyrosynchrotron source at the high-frequency end. 

In all the observed profiles, single $\kappa$ fits of the \fexxiv~255.10~\AA~line perform significantly better than single Gaussian fits. On the other hand, the pronounced wings of the \fexxiv~192.06~\AA~line profiles can in principle be approximated by multiple-Gaussian fits. However, with only a singular exception, such fits are unconstrained. For the \fexxiv~255.1\,\AA~line, this is at least in part due to it red wing occurring outside of the corresponding wavelength windows. Nevertheless, a multiple Gaussian fit gives a width for the dominant \fexxiv~192.06~\AA~component which is unrealistically narrower ($w_\mathrm{G}$ = 0.10$\pm$0.005~\AA~or $FWHM_\mathrm{G}$ = 0.21$\pm$0.04~\AA) than the one obtained in the same location for the other \fexxiv~line at 255.10~\AA~($w_\mathrm{G}$ = 0.13$\pm$0.009~\AA~or $FWHM_\mathrm{G}$ = 0.31$\pm$0.02~\AA).

Considering different possibilities, we show that the observed non-Gaussian line profiles could either be due to local and continuous ion acceleration, or turbulence. This conclusion is in agreement with \citet{Jeffrey17}. The explanation due to multithermal plasma is unlikely, as the equivalent ion temperatures required by the double-Gaussian fits are unrealistically large. Similarly, although the line profiles have similar $\kappa$ as the \emph{RHESSI} high-energy tail, as well as possibly similar temperatures, it is unlikely that the electron beam heats these ions, as the energy content of the beam is too weak.

We note that the $\kappa$ parameter for electrons cannot be diagnosed from line intensities using the ratio-ratio method \citep[cf.,][]{Dudik14a,Dudik15}. We could only use the temperature-sensitive \fexxiv\,/\,\fexxiii~line intensity ratio to diagnose the $T_\mathrm{e}$, if a value of $\kappa$ was assumed. This theoretical temperature-sensitive ratio is shifted to larger $T_\mathrm{e}$ for low $\kappa$, a consequence of the behavior of the ionization equilibrium. Using the observed \fexxiv\,/\,\fexxiii~ratio, we obtain $T_\mathrm{e}$\,=\,27\,MK if a Maxwellian distribution is assumed, while for $\kappa$\,=\,3, the temperature reaches 39\,MK, and for $\kappa$\,=\,2 it reaches 66\,MK. Although this effect leads to an increase of the corresponding thermal velocity for the low $\kappa$ values determined from line profiles or the \emph{RHESSI} high-energy tail, the line profiles are still broad enough to indicate that the approximate lower limit of the non-thermal velocities, if interpreted in terms of turbulence, can be larger than 200\,km\,s$^{-1}$.

We conclude that the non-thermal widths, long observed at the start and impulsive phases of solar flares, are likely connected with extremely non-Gaussian line profiles of the hottest Fe lines, which could be detected here due to good wavelength coverage and resolution of EIS. The present observation of the wide non-Gaussian profiles and HXR/microwave loop-top sources, along with the favorable limb geometry of our flare, provide important observational constraints into the mechanisms responsible for the energy release in the impulsive phase of large solar flares.

\acknowledgments
The authors acknowledge helpful discussions with M. Karlick\'{y}, G. Del Zanna, and H.E. Mason.
J.K. would like to thank the \emph{RHESSI} team, namely K. Tolbert, B. Dennis, S. Krucker and R. Schwartz, for their invaluable advice and software support. V.P.  is supported by NASA grant NNX15AF50G, grant number 2015-065 from the University of Alabama in Huntsville and by contract 8100002705 from Lockheed-Martin to SAO. 
P.T. acknowledges support by NASA grant NNX15AF50G, contract 8100002705 from
Lockheed-Martin to SAO, and contract NNM07AB07C to SAO.

J.D., J.K., and E.Dz. acknowledge support from Grants No. 17-16447S and 18-09072S of the Grant Agency of the Czech Republic, as well as institutional support RVO:67985815 from the Czech Academy of Sciences. B.C. acknowledges support by NASA
grant NNX17AB82G and NSF grant AGS-1654382.
V.P., J.K. and K.K.R. would like to thank the International Space Science Institute (ISSI) for their support and hospitality during the meetings of the ISSI Bern-Bejing team on "Diagnosing heating mechanisms in solar flares through spectroscopic observations of flare ribbons".  
AIA data are courtesy of NASA/\emph{SDO} and the AIA science team. \emph{Hinode} is a Japanese mission developed and launched by ISAS/JAXA, with NAOJ as domestic partner and NASA and STFC (UK) as international partners. It is operated by these agencies in cooperation with ESA and NSC (Norway).
CHIANTI is a collaborative project involving the NRL (USA), the University of Cambridge (UK), and George Mason University (USA).

\bibliographystyle{aasjournal}
\bibliography{kappa}
\end{document}